\documentclass{vldb}
\usepackage{graphicx}
\usepackage{balance}  % for  \balance command ON LAST PAGE  (only there!)

\usepackage{enumitem}
\usepackage{subfigure}
\usepackage{url}

\newcommand{\plotter}{Mr.\ Plotter}

\begin{document}

% ****************** TITLE ****************************************

\title{\plotter{}: Unifying Data Reduction Techniques in Storage and Visualization Systems}

% possible, but not really needed or used for PVLDB:
%\subtitle{[Extended Abstract]
%\titlenote{A full version of this paper is available as\textit{Author's Guide to Preparing ACM SIG Proceedings Using \LaTeX$2_\epsilon$\ and BibTeX} at \texttt{www.acm.org/eaddress.htm}}}

% ****************** AUTHORS **************************************


\numberofauthors{3} %  in this sample file, there are a *total*

\author{
% 1st. author
\alignauthor
Sam Kumar\\
       \affaddr{UC Berkeley}\\
       \email{samkumar@berkeley.edu}
% 2nd. author
\alignauthor
Michael P Andersen\\
       \affaddr{UC Berkeley}\\
       \email{m.andersen@berkeley.edu}
% 3rd. author
\alignauthor
David E. Culler\\
       \affaddr{UC Berkeley}\\
       \email{culler@berkeley.edu}
}
% There's nothing stopping you putting the seventh, eighth, etc.
% author on the opening page (as the 'third row') but we ask,
% for aesthetic reasons that you place these 'additional authors'
% in the \additional authors block, viz.
%\additionalauthors{Additional authors: John Smith (The Th{\o}rv\"{a}ld Group, {\texttt{jsmith@affiliation.org}}), Julius P.~Kumquat
%(The \raggedright{Kumquat} Consortium, {\small \texttt{jpkumquat@consortium.net}}), and Ahmet Sacan (Drexel University, {\small \texttt{ahmetdevel@gmail.com}})}
\date{30 July 1999}
% Just remember to make sure that the TOTAL number of authors
% is the number that will appear on the first page PLUS the
% number that will appear in the \additionalauthors section.

\maketitle

\begin{abstract}
As the rate of data collection continues to grow rapidly, developing visualization tools that scale to immense data sets is a serious and ever-increasing challenge.
Existing approaches generally seek to decouple storage and visualization systems, performing just-in-time data reduction to transparently avoid overloading the visualizer.
We present a new architecture in which the visualizer and data store are tightly coupled. Unlike systems that read raw data from storage, the performance of our system scales linearly with the size of the final visualization, essentially independent of the size of the data.
Thus, it scales to massive data sets while supporting interactive performance (sub-100 ms query latency).
This enables a new class of visualization clients that automatically manage data, quickly and transparently requesting data from the underlying database without requiring the user to explicitly initiate queries.
It lays a groundwork for supporting truly interactive exploration of big data and opens new directions for research on scalable information visualization systems.
\end{abstract}

\section{Introduction}\label{sec:introduction}
As researchers in the 1980s and 1990s were just discovering how to use high-resolution color displays to visualize information, it became clear that interactive visualization and data processing are useful to a wide variety of applications~\cite{shneiderman1994dynamic} and would augment users' ability to perform meaningful analyses~\cite{hellerstein1999interactive}. In 1996, Shneiderman stated his famous information-seeking mantra, ``Overview first, zoom and filter, then details-on-demand''~\cite{shneiderman1996eyes} as a guiding principle in creating data visualization tools. However, this gold standard has proved difficult to realize when working with large datasets. There has been a sharp tradeoff between achieving interactivity and scaling to large datasets, due to the data management problems inherent in creating overview visualizations~\cite{cox1997managing}.

Even as computing and storage technologies have improved vastly over the past two decades, this problem has not abated. In fact, it has worsened. As Hellerstein et al.\ explain in 1999, the appetite for data has grown faster than Moore's Law, meaning that the time to analyze large data sets has steadily grown~\cite{hellerstein1999interactive}, and as Stoica explains in 2013, this trend is only expected to continue~\cite{stoica2013big}. To address this problem, the database and systems communities have developed column-oriented databases~\cite{stonebraker2005c}, parallel databases~\cite{pavlo2009comparison} and distributed computation frameworks~\cite{zaharia2010spark} to process data faster. In addition, researchers have applied sampling techniques~\cite{hellerstein1997online, agarwal2013blinkdb} and precomputation of aggregates~\cite{buevich2013respawn, andersen2016btrdb} to respond to queries at interactive speeds, by providing approximate results or statistical summaries rather than exact results or raw data.

Meanwhile, the data visualization community has arrived at similar solutions. In his keynote paper at SIGMOD 2008, Shneiderman observed that data sets had grown so large that they could no longer be rendered with atomic markers on a single display. He predicted that visualization tools will need to use data reduction techniques, such as filtering, aggregation, or sampling, to solve the problem of ``squeezing a billion records into a million pixels''~\cite{shneiderman2008extreme}.
Recent work in visualization systems makes use of data reduction to scale communication, rendering, and visualization quality to large datasets.
ScalaR~\cite{battle2013dynamic} uses query plan information to decide if data reduction should be used.
M4~\cite{jugel2014m4} renders aggregates instead of raw points.
However, these systems do not take advantage of interactive databases that use similar techniques. They perform queries against traditional databases, performing aggregation just-in-time.

\begin{figure}
    \centering
    \subfigure[A na\"ive design: visualization client requests and renders all points matching a query]{
        \includegraphics[width=\linewidth]{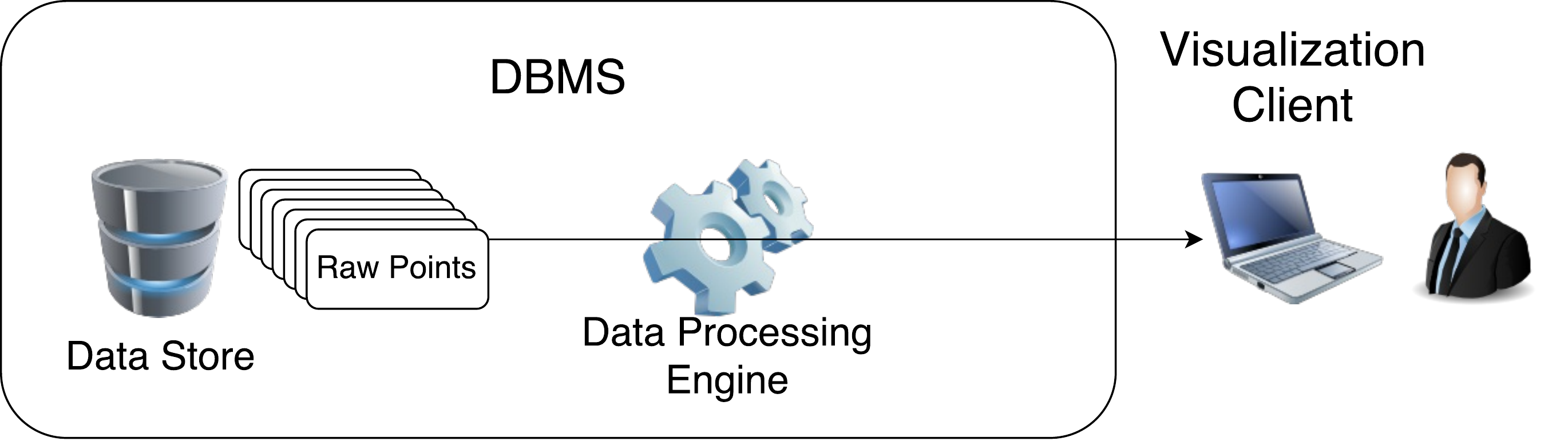}
    }
    \subfigure[A design used by visualization tools like M4 and ScalaR: visualization client requests and renders data aggregates, which are computed by the database system on the fly]{
        \includegraphics[width=\linewidth]{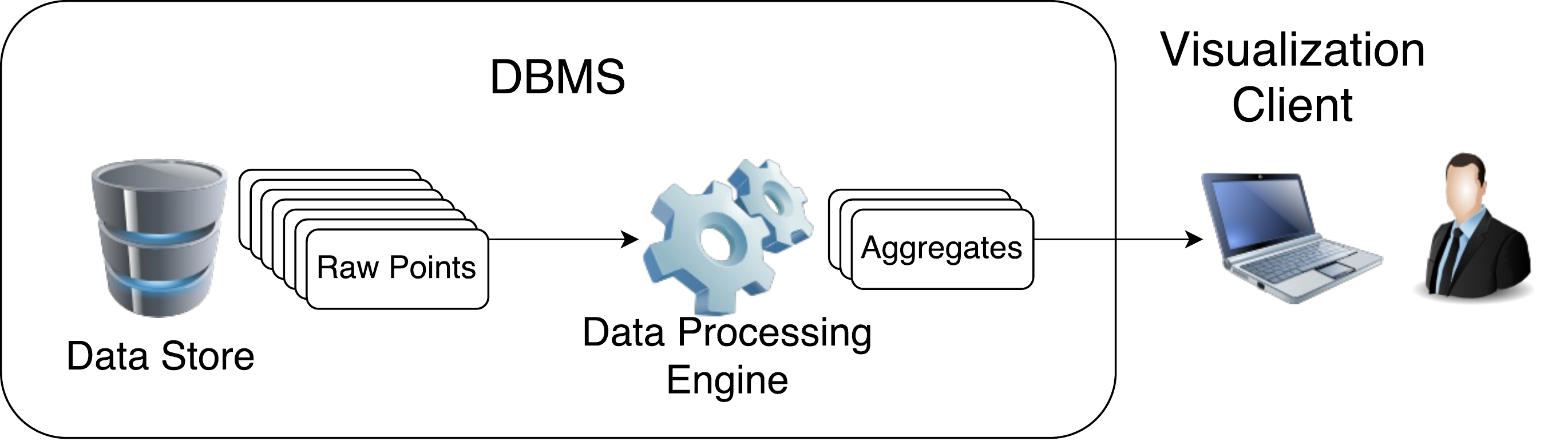}
    }
    \subfigure[\plotter{}'s design: database system accelerates queries using precomputation; visualization client requests and renders precomputed data aggregates]{
        \includegraphics[width=\linewidth]{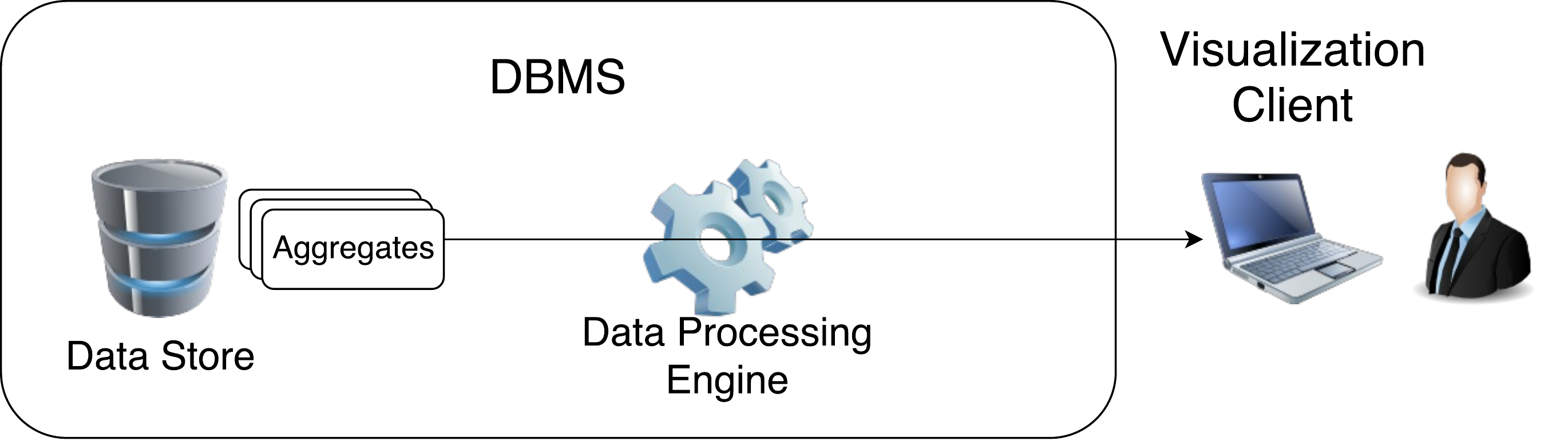}
    }
    \caption{Comparison of approaches to data visualization}
    \label{fig:aggregation_idea}
    \vspace{-12pt}
\end{figure}

In this paper, we present \plotter{}, the \textbf{M}ulti\textbf{r}esolution \textbf{Plotter}, which unifies the data reduction techniques used to create scalable visualizations with those used to create interactive databases. Specifically, we use a data store that uses hierarchical pre-aggregation to provide fast response times, and leverage \emph{the same aggregation technique} to alleviate bottlenecks in rendering, communication, and visual clutter in the visualization client.
As a result, \plotter{} substantially outperforms other visualization tools (see Figure \ref{fig:aggregation_idea} and Table \ref{table:comparison}).

\plotter{} consists of a visualization client and a data store. For the data store, we use the Berkeley Tree Database (BTrDB)~\cite{andersen2016btrdb}, a database for scalar-valued timeseries data that can efficiently serve queries for statistical summaries of data. The visualization client is a custom-built desktop application responsible for providing a user interface and fetching and rendering the appropriate data. By closely coupling these two components, we create a highly responsive and interactive visualization for the user.

Some past work has also tried to closely couple the visualization client and the data store. The prior work most similar to \plotter{} is Skydive~\cite{godfrey2015interactive}, a system for visualizing spatial data. Skydive encodes its data into an \textit{aggregate pyramid}, with raw data at the base and statistical summaries, at progressively larger bin sizes, as one moves up the pyramid. The data visualization client lets the user choose a stratum from the pyramid and a ``cut'' of data (i.e., a working dataset), and then map that data to a visual presentation.

However, \plotter{} improves upon systems like Skydive by incorporating data management into the visualization client.
While the user focuses on zooming and scrolling through the data, the client automatically requests data at the appropriate resolution and time ranges, and shows perceptually relevant information until that data is received from the data store. We call this behavior \textit{automatic data management}.
Furthermore, \plotter{} reconciles client-side caching and automatic data management with the notion of fast data---data that are inserted out-of-order, being streamed into the data store, or otherwise changing. This makes \plotter{} suitable for situations where data are being materialized in real-time~\cite{bailis2017macrobase, andersen2015distil}.

In summary, the main contributions of our work are:
\begin{itemize}[leftmargin=*]
    \item We unify data reduction techniques used by database systems to support interactivity, with those used by visualization systems to eliminate rendering/network bottlenecks and visual clutter.
    \item We demonstrate how to apply hierarchical aggregation to provide interactive visualization of scalar-valued timeseries data.
    \item We provide a visualization client that caches data, automatically and transparently requests data as the user pans and zooms, and shows a perceptually relevant plot until the requests return---behavior that we term \textit{automatic data management}.
    \item We reconcile this with ``fast'' data (i.e., data streaming into the data store, being inserted out-of-order, or otherwise changing).
\end{itemize}

The remainder of this paper is organized as follows. Section \ref{sec:btrdb} provides necessary details about BTrDB, the backend data store used by \plotter{}. Sections \ref{sec:architecture} and \ref{sec:client} describe the design and implementation, respectively, of \plotter{}'s visualization client. Section \ref{sec:evaluation} evaluates \plotter{}. Section \ref{sec:background} compares \plotter{} to related work. Finally, Section \ref{sec:future} discusses directions for future work, and Section \ref{sec:conclusion} concludes the paper.

\begin{table*}
    \centering
    \begin{tabular}{|c||c|c|c|c|c|c|} \hline
        System & ATLAS~\cite{chan2008maintaining} & ScalaR~\cite{battle2013dynamic} & ScalaR~\cite{battle2013dynamic} & M4~\cite{jugel2014m4} & \plotter{} & \plotter{} \\ \hline
        Number of Records & 1.28 Billion & 2.70 Billion & 2.70 Billion & 3.60 Million & 233 Million & 23.3 Billion \\ \hline
        Time to Overview & 28 s & 89.55 s & 1.95 s & 5 s & 223 ms / 56.2 ms & 294 ms / 50.0 ms \\ \hline
        Data Reduction & Aggregation & Aggregation & Sampling & Aggregation & Aggregation & Aggregation \\ \hline
    \end{tabular}
    \caption{Time to produce an overview visualization. We provide two measurements for \plotter{}: (1) latency to fetch aggregates when the server-side in-memory cache is empty, and (2) latency to fetch aggregates with a hot server-side cache.}
    \label{table:comparison}
\end{table*}

\section{Data Storage: BTrDB}\label{sec:btrdb}
This section motivates our choice of aggregation as a data reduction technique, and then explains relevant details about the API and implementation of BTrDB, the data store used by \plotter{}.

\subsection{The Case for Aggregation}\label{ssec:aggregation}

We describe the three main data reduction techniques that database systems and visualization systems use: filtering, sampling, and aggregation. Then, we explain our choice to use aggregation.

\subsubsection{Sampling and Filtering}
In sampling and filtering techniques, a subset of the data is selected, and traditional techniques for query processing or visualization are applied to that subset. In filtering techniques, this subset consists of data that match a certain predicate (e.g., adding a WHERE clause to a SQL query). In sampling techniques, the subset is chosen without attention to a particular predicate on the data; usually, it is chosen randomly.
BlinkDB~\cite{agarwal2013blinkdb} precomputes samples of data, which it uses to produce approximate results to queries within a user-specified time bound or error bound. Visualization tools~\cite{ahlberg1994visual, rafiei2005effectively, park2016visualization} also use these techniques, retrieving only a subset of data from the server to reduce the costs of communication and rendering.

\subsubsection{Aggregation}
In aggregation techniques, the space of data is divided into bins, and aggregate statistics (e.g., number of records, average value, etc.) are computed for each bin. Respawn~\cite{buevich2013respawn} and BTrDB~\cite{andersen2016btrdb} precompute aggregates, accelerating queries for those aggregates. Visualization tools such as ATLAS~\cite{chan2008maintaining}, ScalaR~\cite{battle2013dynamic}, M4~\cite{jugel2014m4}, and Skydive~\cite{godfrey2015interactive} display aggregates instead of raw data. Hierarchical aggregation schemes, in which bins are organized as a tree where each bin is the union of its children, are particularly amenable to visualization \cite{elmqvist2010hierarchical, andersen2015distil, bikakis2017hierarchical}. One can create an overview of a large data set using the aggregate at the root, and drill down to the details by following a branch of the tree, realizing Shneiderman's mantra, ``Overview first, zoom and filter, then details-on-demand.''

\subsubsection{Discussion}\label{sssec:aggregation_discussion}
We use aggregation for data reduction in \plotter{}. The primary reason is that random sampling always has the potential to miss outliers.
In contrast, aggregrates such as ``min'' and ``max'' reliably capture the existence of outliers.
Meanwhile, filtering cannot be used to produce an overview visualization, as the subset is, by construction, not representative of the overall dataset.

Existing research has also investigated non-random samples, chosen so that a sampled plot looks similar to a plot with all of the raw data~\cite{park2016visualization}.
However, such approaches do nothing to remove visual clutter.
In contrast, a plot of aggregates may actually convey \textit{more} useful information than a plot of raw data. With aggregation, one could plot the mean value in each bin; such moving average plots can be more insightful than ones containing raw data~\cite{rong2017asap}, as they are less easily cluttered by local fluctuations than individual data points~\cite{cui2006measuring, elmqvist2010hierarchical}. See Section \ref{ssec:plot} for \plotter{}'s solution.

Other researchers~\cite{liu2013immens} have elected to use aggregation, rather than filtering or sampling, for similar reasons.

\subsection{BTrDB's Timeseries Abstraction}\label{ssec:timeseries}
We use BTrDB~\cite{andersen2016btrdb} for \plotter{}'s data store because it not only supports accelerated aggregate queries, but also provides a unique timeseries abstraction enabling visualization of ``fast data.'' This section describes the parts of BTrDB's API relevant to \plotter{}.

In BTrDB, a \textit{stream} is a representation of a timeseries. A stream is identified by a 128-bit Universally Unique Identifier (UUID), and stores a sequence of (time, value) points. The time is a 64-bit nanosecond timestamp, and the value is a 64-bit floating-point number.
BTrDB is a \textit{copy-on-write} database. When a stream is modified, a new logical version of the stream is created. The old version of the stream still exists in the database and can be queried separately.

The most basic query that BTrDB supports is a query for raw data: \textbf{RawValues(UUID, StartTime, EndTime, Version) $\rightarrow$ (Version, [(Time, Value)])}. This query returns the list of (time, value) pairs in the stream identified by UUID, where StartTime $\leq$ time $<$ EndTime. A special value ``latest,'' that represents the most recent version, can be passed in as the Version argument; in this case, the return value contains the corresponding version number.

The query used most often by \plotter{} is a query for statistical aggregates: \textbf{AlignedWindows(UUID, StartTime, EndTime, Resolution, Version) $\rightarrow$ (Version, [(Time, Min, Mean, Max, Count)])}. To process this query, BTrDB divides the time range [StartTime, EndTime) into intervals that are each $2^\text{Resolution}$ nanoseconds wide. BTrDB then returns the number of points in each time interval, along with the minimum value, mean value, and maximum value of those points.
StartTime and EndTime must be multiples of $2^\text{Resolution}$, hence the name \textit{Aligned} Windows.

\plotter{} makes two more types of queries to BTrDB. The first is \textbf{Nearest(UUID, Time, Direction, Version) $\rightarrow$ (Version, (Time, Value))}, which returns the point immediately before or after the specified time. The Direction argument may be either ``Forward'' or ``Backward,'' specifying in which direction to look for the nearest point. The second is \textbf{Changes(UUID, FromVersion, ToVersion, Resolution) $\rightarrow$ (Version, [(StartTime, EndTime)])}, which returns the time ranges in which two versions of a stream differ. Each returned time range is at least $2^\text{Resolution}$ nanoseconds wide. Thus, the caller can use the Resolution argument to control the granularity at which the time ranges are computed.

BTrDB supports additional queries not described in this paper. For a full description of BTrDB's API, we direct the reader to \cite{andersen2017package}.

\subsection{Summary of BTrDB's Implementation}\label{ssec:hierarchy}
BTrDB supports each type of query in Section \ref{ssec:timeseries} with running time logarithmic in the density of the stream, and linear in the size of the response. In this section we provide a summary of BTrDB's implementation to motivate how it achieves this running time.

BTrDB stores data in a time-partitioning tree. Each stream is represented as a separate tree. The tree is copy-on-write, so whenever a stream is modified, the nodes in the tree that would be modified, and all of their ancestors, are copied; thus each version of a stream corresponds to a separate root node.

The tree can be logically understood as a binary tree. Raw data is stored at the leaves, and each intermediate node stores a statistical aggregate containing the min, mean, max, and count. The tree partitions time. The root node holds the aggregate for the stream over the entire space of time that BTrDB supports. The left child of the root stores the aggregate for the first half of that time range, and the right child stores the aggregate for the second half of that time range. As one travels from root to leaf, the time range is repeatedly divided in half. The nodes at logical depth 63 hold aggregates for each nanosecond, and their children contain the raw data.

When a data point is inserted, the aggregates along the path from leaf to root are recomputed and placed in new nodes, creating a new tree for the new version of the stream. These new nodes may contain pointers to nodes in the old tree. The aggregates (min, mean, max, and count) can be computed at each node using only the node's \textit{immediate} children, making this process very efficient. Note that aggregates are updated at \textit{insertion time}, not query time.

The tree is structured so that the internal nodes of depth $d$ contain statistical aggregates of aligned intervals of size $2^{63-d}$. Therefore, to support the \textbf{AlignedWindows} query, BTrDB simply reads and returns internal nodes of the tree at the correct depth. Furthermore, the tree structure makes satisfying a \textbf{Nearest} query simple: it is easy to avoid empty subtrees. Finally, each node is annotated with the earliest version number it is a part of, making it possible to efficiently satisfy \textbf{Changes} queries: BTrDB traverses the tree corresponding to ToVersion, skipping over subtrees whose root is annotated with a version at least as old as FromVersion, adding a time range to the output whenever this happens.

BTrDB's physical data structure contains important optimizations. First, BTrDB's physical tree has a branching factor of 64, rather than a branching factor of 2. Thus, moving one level up or down the physical tree is equivalent to moving 6 levels in the logical tree.
The intermediate nodes ``missing'' from the physical tree are computed on the fly as they are needed.
Second, BTrDB does not always store raw data at the maximum depth. It may store data higher up in tree, if that subtree represents very few data points. Once the number of data points at a node crosses a threshold, the data are partitioned by time, and a new internal node, storing an aggregate, is created. Third, inserted points are processed in batches, amortizing the cost of recomputing aggregates and creating new nodes. Fourth, BTrDB stores internal nodes in a small storage pool backed by fast Solid-State Drives, and leaf nodes in a large storage pool backed by slower Hard-Disk Drives, taking advantage of the fact that there are many more leaves than internal nodes.

This is just a summary of BTrDB's implementation. For a comprehensive description, see \cite{andersen2016btrdb}.

\section{System Design}\label{sec:architecture}
This section describes the design of \plotter{}, explaining how it supports interactive exploration of scalar-valued timeseries.
Section \ref{ssec:plot} describes the user-facing plot that \plotter{} renders, examining it as a static object. Section \ref{ssec:interaction} describes the plot as a dynamic object, explaining how the user can interact with it to explore timeseries. Section \ref{ssec:automatic_management} discusses data management challenges in supporting this mode of interaction, and our solutions for coping with them.

\subsection{User-Facing Plot}\label{ssec:plot}
Central to Shneiderman's problem of ``squeezing a billion records into a million pixels''~\cite{shneiderman2008extreme} is drawing plots where multiple points map to the same pixel. Aggregation techniques are a natural solution: where multiple atomic markers would ``collide'' in the same pixel, a single aggregate marker can be used. In \plotter{}, we consider line plots with time on the horizontal axis, and value on the vertical axis. Our primary consideration is densely-packed timeseries, where adjacent points may map to the same pixel column.

Solutions such as M4~\cite{jugel2014m4} compute data aggregates to produce a plot that is equivalent, pixel-for-pixel, to a plot drawn with the raw data points. While such a solution eliminates communication and rendering bottlenecks, it does not solve the problem of visual clutter. Plots of raw data points, especially over long periods of time, are often too cluttered for meaningful analysis, because short-term fluctuations tend to hide long-term trends~\cite{cui2006measuring, elmqvist2010hierarchical, rong2017asap}.

\begin{figure}
    \centering
    \subfigure[A plot of a single stream, rendered by \plotter{}]{
        \includegraphics[width=\linewidth]{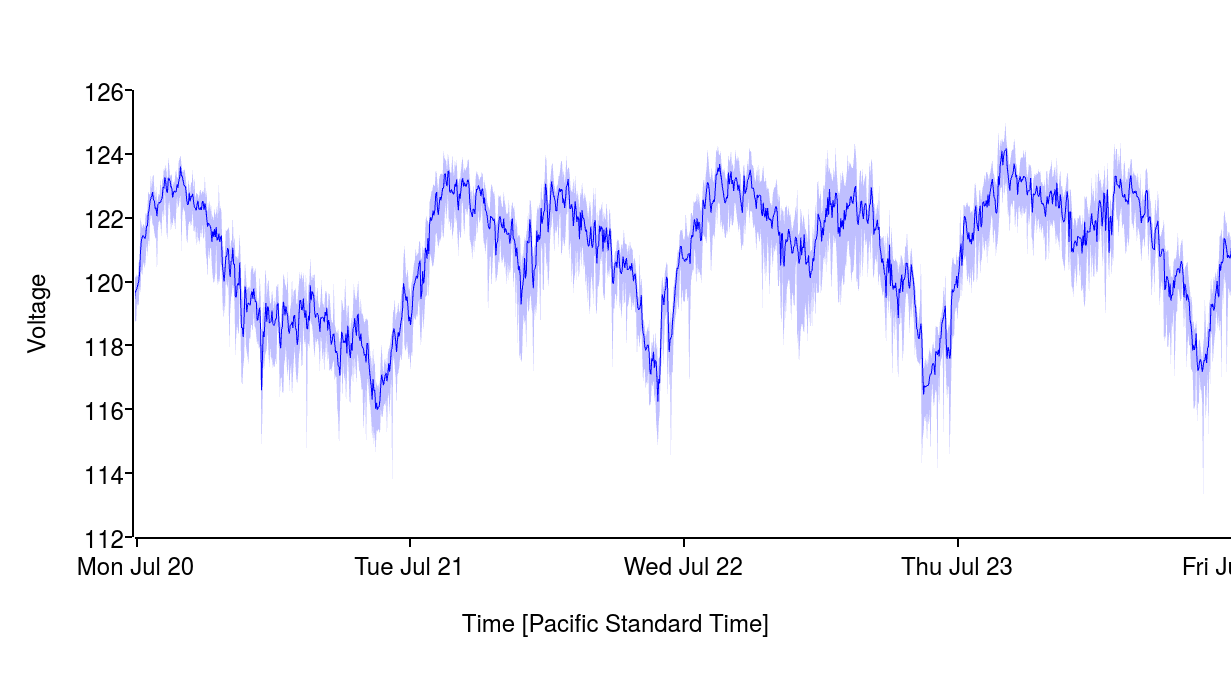}
        \label{subfig:disp_normal}
    }
    \subfigure[Same data, raw points rendered and connected with lines]{
        \includegraphics[width=\linewidth]{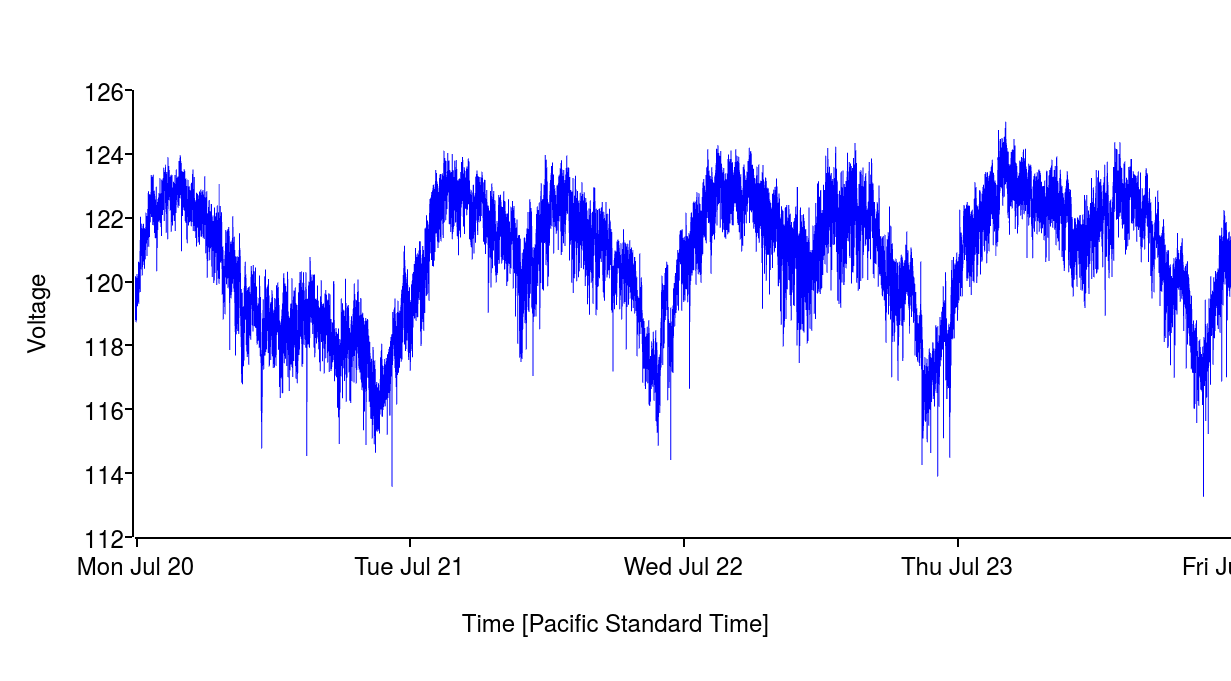}
        \label{subfig:disp_raw_lines}
    }
    \caption{Comparison of a plot of aggregates, as drawn by \plotter{}, to a plot of raw data points}
    \label{fig:aggregation_plot}
    \vspace{-12pt}
\end{figure}

\plotter{} uses a different approach. It uses the ``min'' and ``max'' values of aggregates to draw a translucent silhouette of what the plot would look like if all of the raw points were plotted, and then draws a line plotting the ``mean'' values over the same period. This shows a smoothed plot of mean values that is less cluttered, while still indicating the variance in the data pre-smoothing, via the min-max silhouette. See Figure \ref{subfig:disp_normal} for an example. Because the min-max silhouette is translucent, it does not block large areas of the plot, even if the stream is very noisy. This makes plots with multiple streams less cluttered.
Our plot provides strictly more information than a plot of raw data (Figure \ref{subfig:disp_raw_lines}), while reducing visual clutter.

To indicate missing data, \plotter{} draws a data density plot when a user selects a stream. This displays the ``count" aggregates for the selected stream (see Figure \ref{fig:cases}).
We also emphasize missing data in the main plot by rendering gaps in the data as gaps in the visualization\footnote{Optionally, the user may turn off this behavior, to always connect neighboring points.}. Neighboring aggregates are drawn connected to each other as long as they are temporally adjacent\footnote{BTrDB omits aggregates with $\text{count}=0$ when responding to \textbf{AlignedWindows} queries, in order to save bandwidth. The client infers the existence of gaps when the aggregates returned by BTrDB are not temporally adjacent.}.
This implies that, if the user has zoomed in so far that there is a gap between every pair of neighboring points, \plotter{} will no longer connect the points, effectively displaying a scatterplot.

\subsection{Interaction with the Plot}\label{ssec:interaction}
We support a very fluid mode of interaction with the plot. The user clicks and drags the plot in order to scroll horizontally, and uses her mousewheel or touchpad to zoom in or out. The user does not manually initiate queries to fetch new data; the visualization client automatically queries data depending on the user's current view.

The success of this interaction model depends on timely processing of queries by the database; data must be loaded at interactive, or near-interactive, speeds. We make this possible by unifying the aggregation technique BTrDB uses to accelerate queries, with those that the visualization client uses for rendering.

Even so, we cannot depend on requests to the database to always return at interactive speeds. First, the load on the server could vary, causing query processing times to increase when the server load is heavy. Second, network characteristics may affect performance. Depending on the physical location of the client and server, the network round-trip time may exceed 100 ms.

\begin{figure}
    \centering
    \subfigure[User starts at this view, and then rapidly zooms in to the data near December 18]{
        \includegraphics[width=\linewidth]{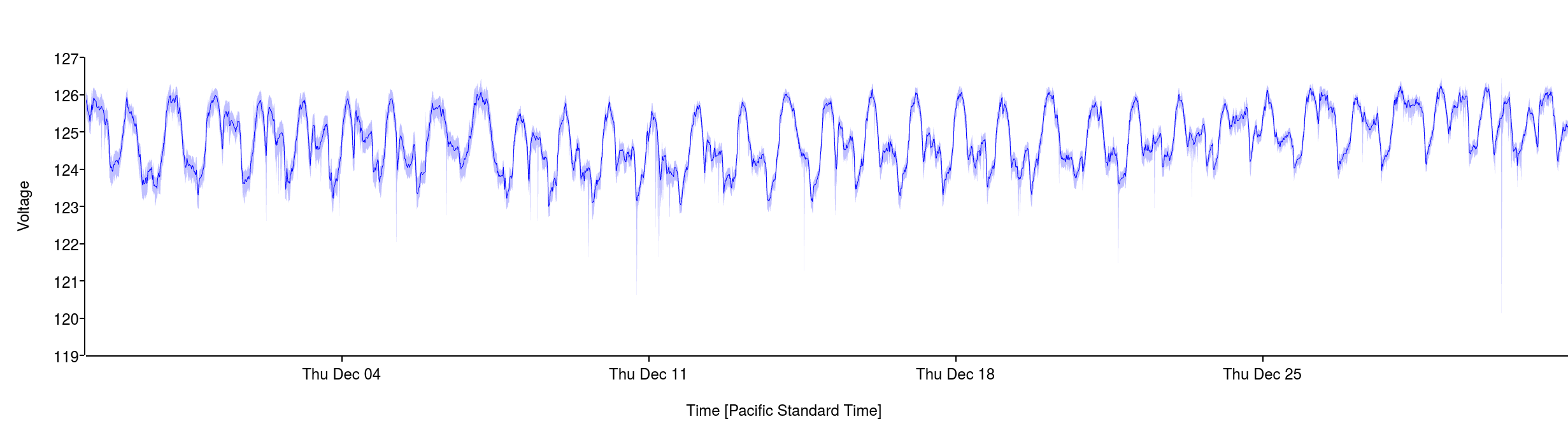}
        \label{subfig:zoom_before}
    }
    \subfigure[\plotter{} continues displaying aggregates from the previous view, transformed to the new axes, while waiting for BTrDB to return data at the new resolution]{
        \includegraphics[width=\linewidth]{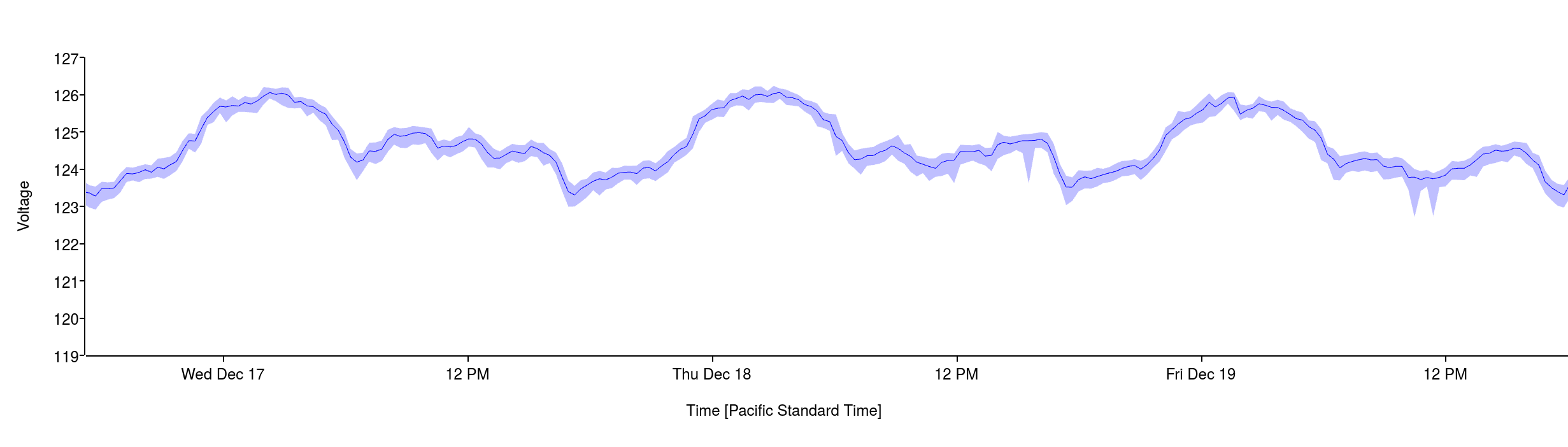}
        \label{subfig:zoom_loading}
    }
    \subfigure[When data at the new resolution finally arrives (usually within 100 milliseconds), the plot is updated]{
        \includegraphics[width=\linewidth]{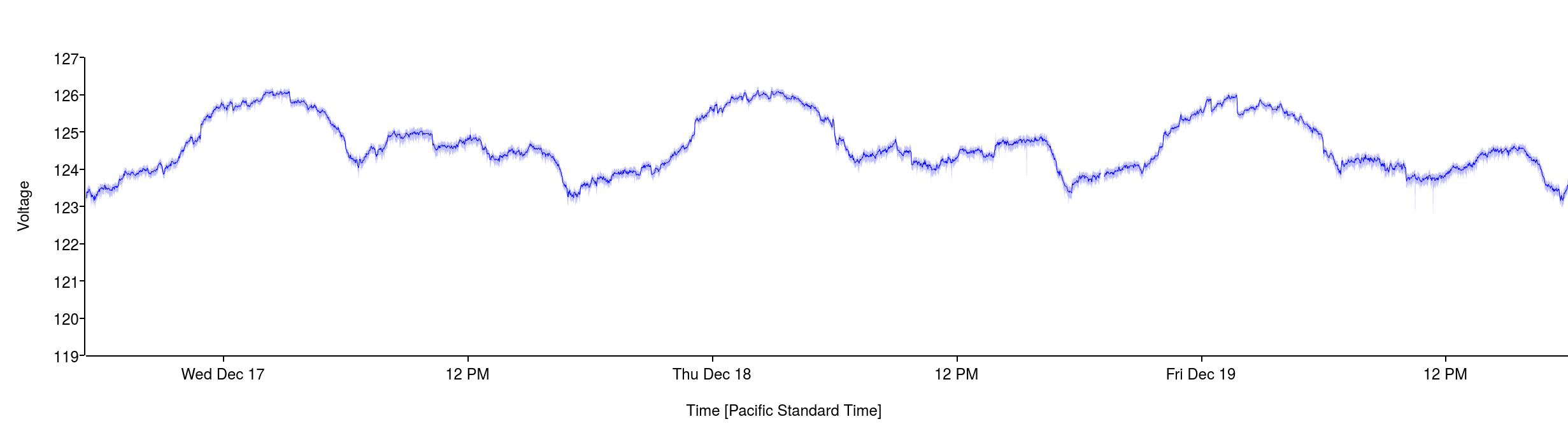}
        \label{subfig:zoom_after}
    }
    \caption{When the desired data is not available on time, \plotter{} continues to display perceptually relevant data}
    \label{fig:zoom}
    \vspace{-12pt}
\end{figure}

The visualization client manages queries to the BTrDB server and responses from the server, to mitigate the impact if some requests take longer to return.
First, whenever aggregates are received from the server, they are placed in an in-memory cache, decreasing the average wait time for new aggregates to load.
Second, when the user zooms or scrolls to a region whose data is not in the cache, \plotter{} continues showing aggregates in the previous view, transformed according to the current view, until the new aggregates are received. In this way, \plotter{} shows perceptually relevant information, even if the ``ideal'' data is not yet available. See Figure \ref{fig:zoom} for an example.
Finally, data is pre-fetched according to the user's current view, further reducing the user's expected wait time. Collectively, we call this behavior \textit{automatic data management}.

\subsection{Supporting Automatic Data Management}\label{ssec:automatic_management}

This section explains aspects of \plotter{}'s design that support the mode of interaction described above.

\subsubsection{Size and Alignment of Aggregates}\label{sssec:alignment}
Solutions such as M4~\cite{jugel2014m4} compute an aggregate of data for each pixel column of the plot, and display each aggregate in its pixel column. The rationale is that rendering multiple aggregates per pixel column results in \textit{implicit} loss of detail.
Therefore, requesting finer-grained aggregates from the database wastes network bandwidth and rendering time.

However, requiring each aggregate to correspond exactly to one pixel column is very restrictive. If the user chooses to zoom in by a small amount, the size of the time interval represented by each pixel column decreases slightly, requiring new aggregates to be computed. Similarly, if the user's view shifts to the left or right, the alignment of time intervals corresponding of pixel columns may change, requiring new aggregates to be computed.

This model introduces nontrivial problems when working with client-side caching and precomputed aggregates. First, caching aggregates returned by the server is only useful for drawing other plots at the \textit{exact same zoom}, and at the \textit{exact same alignment} between time and pixels, as the plot for which the aggregates were requested. Every time the user changes zoom, an additional request for the entire screen of data would have to be made to the server. Second, the client must request aggregates for time intervals whose size and alignment can vary arbitrarily.
BTrDB can quickly return aggregates over time intervals whose endpoints, in nanoseconds, are aligned to multiples of a power of two; however, pixel columns on the user's screen need not be aligned in any particular way\footnote{BTrDB does support \textbf{Windows} queries that compute aggregates over unaligned, arbitrarily-sized intervals, but processing these queries is less efficient than processing \textbf{AlignedWindows} queries. See \cite{andersen2017package} for details.}.

\plotter{} relaxes this constraint. Rather than requiring aggregates to correspond exactly to pixel columns, \plotter{} only requires that each aggregate represent a time interval that is \textit{at most} one pixel column in width. To choose the size of aggregates, \plotter{} takes the size of a pixel column in nanoseconds---the size of aggregates that M4 uses---and rounds it down to the nearest power of two.
This is always a power of two, so the aggregates can be efficiently obtained from BTrDB. Because we no longer require the aggregates to correspond directly to pixel columns, there are no alignment issues\footnote{Aggregates can be rendered even if they span multiple pixel columns, or if a pixel column contain multiple aggregates; this is handled in the graphics pipeline, taking advantage of multisample anti-aliasing if the client's graphics card supports it.}.
Thus, the client can effectively cache aggregates obtained from the server: the aggregates used to draw one plot can be used to draw another overlapping plot, even if the zoom and translation are slightly different.

One may argue that this approach is somewhat wasteful: more aggregates are drawn than pixel columns. We can compute an upper bound on how many extra aggregates are rendered. The size of aggregates, computed as per above, has the following property:
$$\text{Pixel column size} \geq \text{Aggregate size} > \frac{1}{2}\cdot\text{Pixel column size}$$
where the sizes of pixel columns and aggregates are measured in time. We can then divide the width of the screen (in time) by each of these three quantities to show that:
$$\text{Pixel columns} \leq \text{Aggregates rendered} < 2\cdot\text{Pixel columns}$$
which means that we are overplotting by less than a factor of two.

\subsubsection{Prefetching Policy}
As described in Section \ref{ssec:interaction}, \plotter{} prefetches data to hide the latency of requesting and receiving aggregates from the database.
\plotter{} uses a stateless prefetching scheme. At any given view, there are four actions that the user can take: (1) scroll to the left, (2) scroll to the right, (3) zoom in, or (4) zoom out. Once \plotter{} has received the data for the current view, it prefetches data, assuming that the user may take any of these actions. Specifically, \plotter{} prefetches:
\begin{itemize}[leftmargin=*]
    \item One screen of data to the left of the current view
    \item One screen of data to the right of the current view
    \item The current screen's data at the next smaller aggregate size
    \item The current screen's data, and the two neighboring screens' data, at the next larger aggregate size
\end{itemize}
The purpose of the first two items is to prefetch a scrolling gesture. We prefetch one screen's worth of data to the left and right, because of the click-and-drag interface that \plotter{} supports; the user is not likely to move the view by more than one screen width in one gesture. The purpose of the last two items is to prefetch a zooming gesture. The user uses the mousewheel to zoom, and the cursor's current position acts as the focus of the operation. The user can zoom in at any point of the screen, so we prefetch all of the current screen's data at the next smaller size of aggregates. When the user zooms out with the cursor (focus) on one side of the screen, most of the extra time range comes from the other side of the screen; furthermore, if the user zooms out to the next larger aggregate size, the total time range is doubled, because consecutive aggregate sizes vary by factors of two. Therefore, we prefetch the two neighboring screens at the next larger size of aggregates.

Prefetching is not a new idea. For example, ATLAS~\cite{chan2008maintaining} is a visualization tool for scalar-valued timeseries that uses prefetching.
Like \plotter{}, its ``objective is not to pull the maximum amount of data possible, but rather to get the minimum amount of data to sustain smooth interactions.'' To perform prefetching effectively, ATLAS maintains a timing model of the database, allowing it to estimate the latency of each query.
It limits how fast the user can scroll, to ensure that prefetched data will arrive in time.

Although ATLAS' prefetching policy is more sophisticated than \plotter{}'s, a prefetching methodology similar to ATLAS' would not be very useful in \plotter{} due to some key differences between the two systems. First, \plotter{}, unlike ATLAS, does not restrict how fast the user can move between views. Although \plotter{} uses mouse input, which has its limits (it is humanly possible only to move the mouse so fast), it is difficult to characterize how fast the user can move between views. As a result, it is less meaningful to schedule prefetch requests to meet deadlines based on how fast the user can zoom or scroll. Second, and more importantly, ATLAS' prefetching methodology rests on the assumption that ``query processing times increase linearly with the number of records contained in the query range.'' This assumption does not hold for \plotter{}. Due to the hierarchical pre-aggregation scheme described in Section \ref{ssec:hierarchy}, the time taken by BTrDB to process an \textbf{AlignedWindows} query is linear in the number of aggregates returned, not in the number of raw records those aggregates comprise. For most modern screens, which are at most 3000 or 4000 pixels wide, results from \cite{andersen2016btrdb} (confirmed in Section \ref{sec:evaluation}) suggest that query processing in BTrDB will take at most a few hundred milliseconds---barely enough time for a user to click and drag the plot across the screen. Thus, prefetching only adjacent screens is sufficient.

There exist proposals for dynamic prefetching based on similar assumptions to \plotter{} (e.g., ForeCache~\cite{battle2016dynamic}). However, our experience is that \plotter{}'s simple prefetching strategy performs well (see Section \ref{sec:evaluation}). Therefore, we leave an investigation of more sophisticated prefetching techniques to future work.

\subsubsection{Request Throttling}\label{sssec:throttling}
An advantage of plotting the ``min'' and ``max'' is that anomalies are visible even in overview visualizations. If a user notices an anomaly, she may rapidly zoom in to examine it in more detail. As the user zooms in, the plot moves through many levels of resolution. The user does not stop at each resolution before continuing to zoom in, so the ``think time'' between individual gestures is very low\footnote{Because \plotter{} continues to show perceptually relevant information until the desired data is available (see Section \ref{ssec:interaction}), the user is able to continue zooming.}.

A na\"ive implementation would perform a full data refresh on every frame. However, this is wasteful, because the user zooms to the next resolution before each request for data returns. Furthermore, after the user stops zooming in, the request for the final plot would be placed behind these extraneous requests on the server-side request queue, degrading the user's experience.

To solve this problem, \plotter{} implements request throttling. If a cache miss has occurred and a request sent to BTrDB in the past 300 ms, cache lookups are done without making requests to BTrDB on misses; the on-screen data set is left unchanged. \plotter{} smoothly updates the current view as the user zooms and scrolls, using only the on-screen data set, so the plot continues to feel responsive. This limits the number of extraneous requests.

\section{Visualization Client}\label{sec:client}

\begin{figure}
    \centering
    \includegraphics[width=\linewidth]{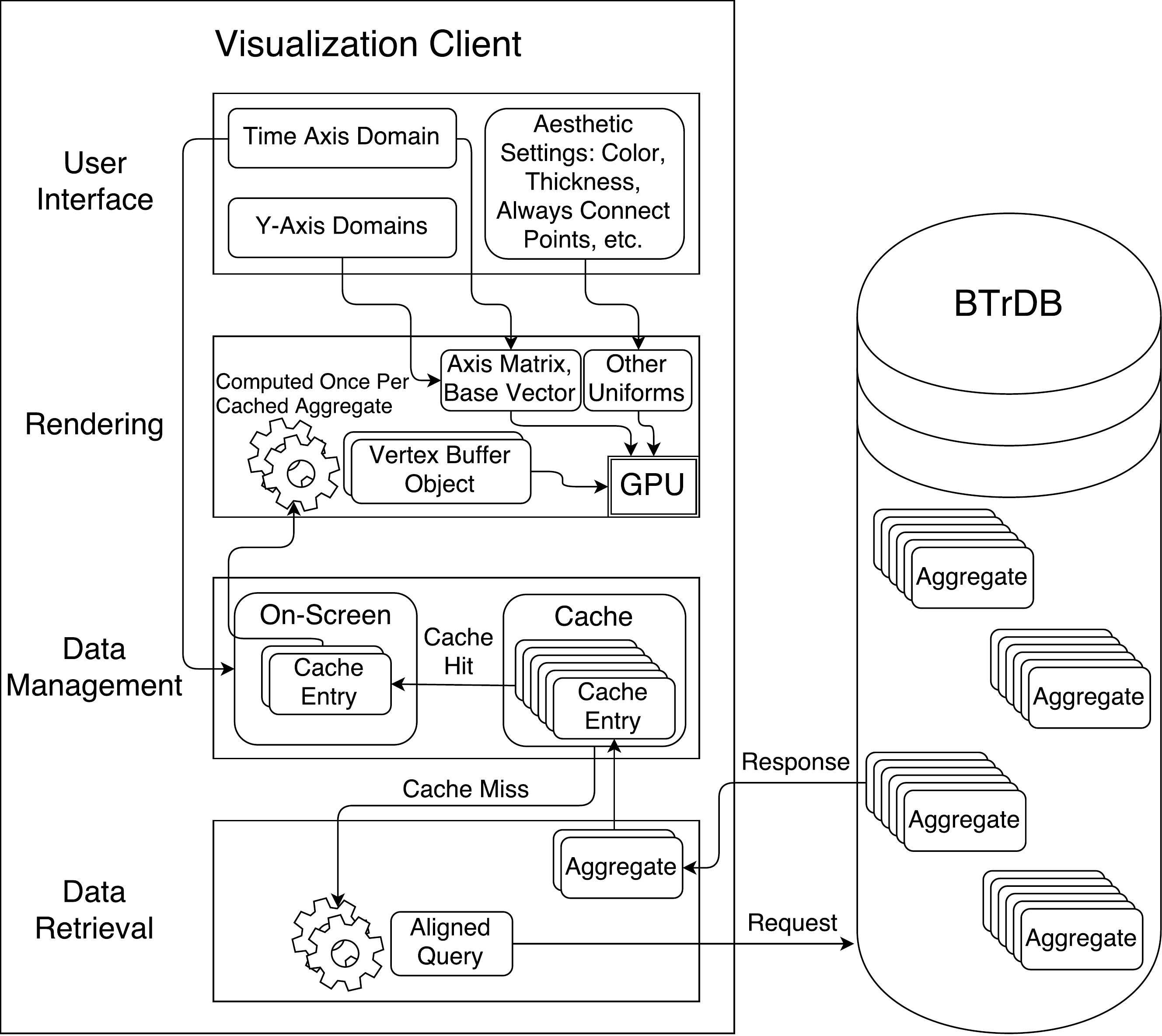}
    \caption{System architecture of \plotter{}}
    \label{fig:architecture}
\end{figure}

This section describes the implementation of the visualization client. The visualization client is a desktop application written in the Qt framework. It is written primarily in C++, but parts of the user interface are written in QML.
As depicted in Figure \ref{fig:architecture}, the visualization client has four layers: Data Retrieval, Data Management, Rendering, and User Interface. We describe each layer in turn.

\subsection{Data Retrieval Layer}
The Data Retrieval Layer is requests data from BTrDB. On cache misses, the Data Retrieval Layer requests aggregates from BTrDB by making \textbf{AlignedWindows} queries. It takes care to properly align queries to multiples of the appropriate powers of two, and leverages gzip compression to make more efficient use of network bandwidth.

\subsection{Data Management Layer}
The Data Management Layer issues queries to the Data Retrieval Layer, maintains a cache of the resulting aggregates, and prefetches data according to the user's current view.

\subsubsection{Cache Organization}\label{sssec:cache}
The in-memory cache stores aggregates obtained from the server in past queries to prevent additional queries for the same data, increasing \plotter{}'s responsiveness.

When the user navigates to a certain view, \plotter{} computes the appropriate resolution (size of aggregates), and then requests aggregates at that resolution for each visible stream, in the time range that is visible.
We use a tree-based map to store the aggregates for each stream and resolution, to allow for efficient lookup of aggregates in a time range.
We organize these trees in array-based map data structures, to allow efficient lookup of the tree of aggregates given a specific stream and resolution. We keep track of the cache size, and use the LRU (``Least Recently Used'') cache eviction policy to keep it within a strict upper bound.

However, storing the aggregates directly in a tree structure is inefficient.
Aggregates must be individually tagged for LRU cache management, adding additional overhead.
Information regarding gaps in the data is also lost in this scheme; the absence of an aggregate in the cache could mean that there is no data for that time range, or it could mean that the aggregate has not been fetched yet.

Our solution is to manage aggregates in chunks, instead of individually. The cache holds data as a set of cache entries. A \textit{cache entry} contains a set of aggregates for a specific stream at a specific resolution, received in response to a single \textbf{AlignedWindows} query, stored as an array. It also contains the start and end timestamps for the query that was performed.
Cache entries keep track of gaps in the data; the existence of a cache entry guarantees that \textit{all} of the data in the specified time range, for the specified stream and resolution, is in the cache entry.
The cache stores and manipulates cache entries as single units, for lookup and LRU eviction. This helps amortize the overhead of storing many aggregates. Furthermore, it makes it easier to prepare aggregates for rendering, because data is passed to the GPU as arrays (see Section \ref{ssec:rendering}).

\subsubsection{Fast Data}
To handle changing data, we maintain a \textit{base version number} for each stream. The base version number is the most recent version such that all of the cache entries for that stream were obtained from a version of the stream at least as recent as the base version. Stated differently, the base version number is a bound on how stale the cached data for a stream can be.

When we make an \textbf{AlignedWindows} query to BTrDB, we receive both the aggregates and the version number of the stream from which those aggregates were obtained. When the cache receives the aggregates, it sets the base version number to the minimum of the current base version number, and the version number of the received data. We periodically make \textbf{Changes} queries to BTrDB, querying the changes between the base version and the latest version. When we receive the time ranges containing the differences, we drop all cache entries containing data in those time ranges, and update the base version to the latest version of the stream, as returned by the \textbf{Changes} query.

\subsubsection{On-Screen Data}
One of the functions of the Data Management Layer is to continue displaying perceptually relevant information when requests miss in the cache and are sent to BTrDB. As described in Section \ref{sec:architecture}, \plotter{} continues to display the aggregates last displayed on the screen, mapped to the current view, while data is loading.
To implement this, the Data Management Layer explicitly manages a collection of cache entries that are rendered on-screen.
When the view changes, the data for the new time range and resolution are looked up in the cache, and missing aggregates are requested from the server. Until the new data arrive, the on-screen data are not updated, and therefore continue to be mapped to the current view and rendered.

\subsubsection{Cache Misses}
When the user updates her view, \plotter{} looks in the cache to update the on-screen data set. When not all of the required data is in the cache, a request is sent to the Data Retrieval Layer, which makes a request to BTrDB.

However, even before the data is loaded, the user may slightly adjust their plot. In such a scenario, we should not repeat the request to BTrDB. Instead, we should wait for the current outstanding request to finish, and only request data that will not be returned by the request we have already made. To do this, we create a placeholder cache entry with a ``pending'' bit set to indicate that a request for that data is outstanding. When another cache miss occurs for that data, the caller sees that there is already a pending request, and waits for the response instead of issuing a new request.

\subsection{Rendering Layer}\label{ssec:rendering}
The Rendering Layer is responsible for converting the on-screen data set, maintained by the Data Management Layer, into a user-facing plot. To render the plot, we use OpenGL, an API for interacting with a GPU for hardware-accelerated graphics rendering.

The first time a cache entry is rendered, it is converted to a \textit{Vertex Buffer Object}, or VBO. A VBO is a buffer of data in graphics memory, ready for processing by a shader program running on the GPU. The VBO handle is stored in the cache entry, for easy retrieval. Then, the VBOs for cache entries in the on-screen data set, along with
the domains of the time and value axes and aesthetic settings chosen by the user, are passed to a shader program, which renders the plot. %
This section describes the rendering pipeline in detail.

\subsubsection{Rendering an Aggregate}
Before discussing how to generate VBOs from cache entries, we discuss how aggregates are rendered. %
The ``min,'' ``mean,'' and ``max'' values are drawn at the midpoint of the time interval represented by that aggregate.
If the neighboring aggregates are temporally adjacent (i.e., there is no gap in between), then the aggregates are connected: the mean values are connected by a line, and the min-max values are connected by a translucent trapezoid.
If this aggregate is not temporally adjacent to either of its neighbors, then a translucent vertical line is drawn connecting the min and max, and the mean is depicted as a single point (as in a scatterplot). This often happens when the aggregate contains a single raw data point, in which case the min and max are equal and the vertical line is not visible.

\begin{figure}
    \centering
    \subfigure[The first three aggregates have gaps on both sides and are rendered as lone points, whereas the remaining six are temporally adjacent and are rendered connected]{
        \includegraphics[width=\linewidth]{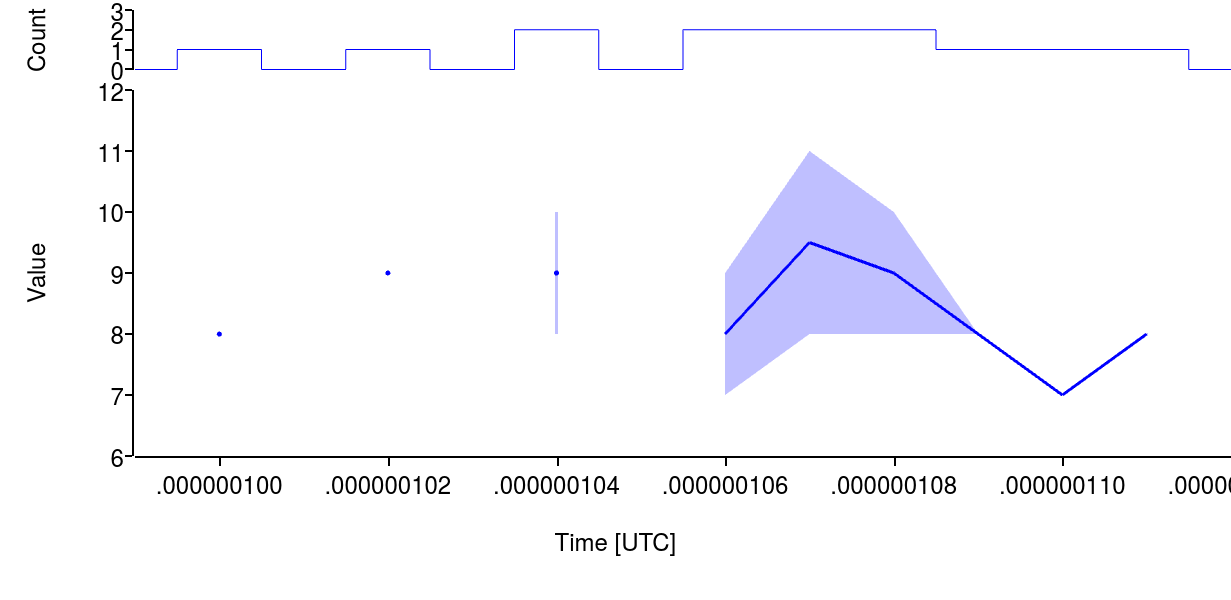}
        \label{subfig:cases_default}
    }
    \subfigure[Same plot as Figure \ref{subfig:cases_default}, with the ``Always connect points'' option selected]{
        \includegraphics[width=\linewidth]{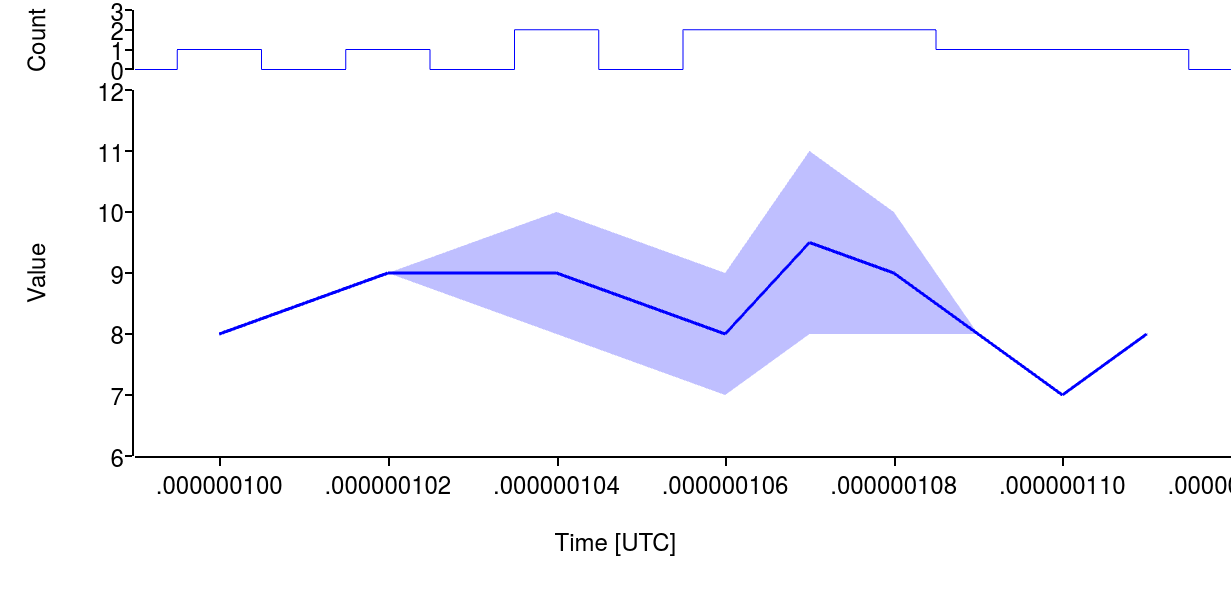}
        \label{subfig:cases_connected}
    }
    \caption{Plots that display all cases for rendering aggregates}
    \label{fig:cases}
    \vspace{-12pt}
\end{figure}

Figure \ref{fig:cases} demonstrates how aggregates are rendered. Note that temporally adjacent aggregates are normally at most one pixel column apart; these plots are zoomed in beyond the granularity of nanosecond timestamps, to emphasize individual aggregates.

\subsubsection{Generation of Vertex Buffer Objects}\label{sssec:vbos}
To draw the min-max translucent silhouette, we need to pass the min and max for each aggregate as separate vertices to the GPU.
Therefore, we generate two vertices in the VBO for every aggregate in the cache entry.
Then, we render the data as a triangle strip.

To ensure that gaps are rendered properly, we must include information in the VBOs that the shader program can use to selectively omit certain triangles. Our approach is to insert two new points in every gap, with a special flag value indicating that their purpose is to mark a gap; the vertex shader sets a ``do not render'' flag for these points. The value of the ``do not render'' flag is interpolated between all of the pixels; we discard fragments where the interpolated ``do not render flag'' is strictly greater than 0, thereby drawing a gap. %

To draw the mean line, we make a second pass, rendering a line strip instead of a triangle strip. We also use only one point per aggregate instead of two. We then make two more rendering passes, drawing vertical lines and points, to render lone points.
When generating the VBOs, we identify lone points and set special flags, so that only those points are rendered during these passes.

Finally, the user has an option to ``Always connect points.'' With this option enabled, neighboring points are always connected, regardless of whether they are temporally adjacent. To implement this, we pass this option to the vertex shader as another uniform variable; if the option is set, then the ``do not render'' flag is never set. Note that the vertices do not change, so the vertices in each gap are still present. However, we can position them appropriately so that they do not interfere with the final image. See Figure \ref{subfig:cases_connected}.

\subsubsection{Data Density Plot}\label{sssec:density}
As described in Section \ref{ssec:plot}, a data density plot is drawn when the user selects a stream. This plot displays the ``count'' aggregates, pulling the plot down to 0 where aggregates are missing. To pull the plot to 0, we insert an extra pair of points in each gap, at the timestamp where the the first aggregate in the gap would be if it were present. Two points are needed so that the count maintains a step-like appearance.
Because we need two extra points anyway to render the gap in the min-max silhouette (Section \ref{sssec:vbos}), this comes at no additional cost.

\subsubsection{Shader Programs}
The shader programs map the generated vertices, passed in via a VBO, to pixels on the screen. The user's current view is passed to the GPU as a base vector $\vec{v}_\text{base}\in\mathbb{R}^2$ containing the (time, value) coordinates of the lower left-hand coordinate of the screen, and an axis matrix $A\in M_3(\mathbb{R})$ based on the sizes of the time and value axes. To map a (time, value) coordinate $\vec{x}$ to its final position, we compute $A\cdot\text{pad}(\vec{x} - \vec{v}_\text{base})$, where ``pad'' converts a vector in $\mathbb{R}^2$ to a vector in $\mathbb{R}^3$ by padding it with a 1.

One subtlety is that this transformation applies to floating point numbers, but the timestamp is stored as a 64-bit integer representing the number of nanoseconds since 1970. Converting this to a single-precision floating point number would result in an unacceptable loss of precision. To solve this problem, we store timestamps relative to an \textit{epoch time}, which is computed as the midpoint of the cache entry. Any differences between timestamps in the cache entry and the epoch time will be on the same order as the full time range currently displayed. Therefore, the difference between the timestamps in the cache entry and the epoch time can be safely converted to a single-precision floating point number without losing too much precision.
The base vector is also computed relative to the epoch time.

\subsubsection{Additional Considerations}
In the rendering pipeline, each cache entry is drawn independently. However, the aggregate at the end of one cache entry may be temporally adjacent to the aggregate at the beginning of the next cache entry, meaning that the two points should be connected. As the points will be in separate VBOs, this is not possible. To solve this problem, we modify each query to include the aggregate immediately before and the aggregate immediately after the queried time interval. When a cache entry is created, it checks if the cache entries immediately before it and immediately after it have been created and connect to its ``edge'' points; if not, it takes ownership of connecting to those points.

Another issue is that the CPU and GPU must communicate to initiate the rendering of each cache entry. This overhead may become significant if many small cache entries must be drawn for a certain view. Our solution is to avoid making queries for tiny segments of data, to prevent cache entries from becoming too fragmented. If the user scrolls by a small amount, we request the next screen width's worth of data, rather than just the small interval that is newly visible. This guarantees that \emph{at most five cache entries must be drawn to render any view of a single stream}, which is a good enough upper bound to maintain qualitatively good performance.

\subsection{User Interface Layer}
The User Interface Layer provides the click-and-drag interface for scrolling and the mousewheel interface for zooming. In response to click-and-drag and mousewheel events, the User Interface Layer updates the user's current view of the plot.
Because the user interface handles user input and sees individual gestures, it is a natural place to handle request throttling.

\section{Evaluation}\label{sec:evaluation}
This section evaluates the performance of \plotter{} and quantifies the extent to which it supports interactive exploration of big data.

\subsection{Performance Model}\label{ssec:model}
Our primary goal is to support interactive exploration of data, so our main metric is the time to access data. Drawing from models of the memory hierarchy of a computer system, we model this as $\overline{\text{Data Access Time}} = \text{Hit Time} + \text{Miss Rate}\times \text{Miss Penalty}$.
Hit Time is the time to access data in the client-side in-memory cache. Miss Rate is the fraction of requests that miss in the cache. Miss Penalty is the time to request data from BTrDB after a cache miss.

Prefetching issues requests for data preemptively. It reduces the miss rate by eliminating compulsory misses, and reduces the miss penalty by issuing requests for data early.

Unlike memory accesses in a computer system, data accesses in \plotter{} happen at perceptible time scales and are individually visible to the user.
Therefore, the \textit{worst-case} access time, $\text{Hit Time} + \text{Miss Penalty}$, is also important.

\subsection{Experimental Methodology}\label{ssec:workload}
Our workload for characterizing \plotter{} is based on Shneiderman's information-seeking mantra: ``Overview first, zoom and filter, then details-on-demand''~\cite{shneiderman1996eyes}. The user begins with an overview visualization that shows all of the data for a stream. Then the user picks an interesting region of the plot, and zooms in to investigate. The user may again identify an interesting region of the zoomed-in plot, and zoom in further, and repeat the process until she is looking at individual points (the ``details'').

We use a Python script to simulate this use case programmatically. It chooses a random point on the time axis of the plot. Then the script programmatically moves the mouse to that point (which takes 100 ms), and zooms in by a fixed amount. The script repeats this process until individual points are visible. In between iterations, it pauses, to simulate the user's ``think time'' as she selects an interesting point to zoom in.
We repeat the experiment multiple times, varying the think time, to measure how well \plotter{} performs. For the data stream that we used, it took 16 iterations to zoom in from an overview visualization comprising 233 million points over three weeks, to individual points spaced 120 ms apart.

We ran BTrDB on a server with an Intel Xeon E5 2640 v4 @ 2.40 GHz and network-attached storage. BTrDB was configured to use no more than 64 GiB of memory. The \plotter{} client ran on a laptop with an Intel Core i7-7820HQ @ 2.90 GHz, and was configured to use no more than 1 GiB of memory to cache aggregates from BTrDB. The client and server were separated by one network router, with a network round-trip time of approximately 0.5 ms.

\subsection{Hit Time}
In each simulation that we performed using the above workload, we measure the cache lookup time, by obtaining a millisecond-precise timestamp before each cache lookup and after the requested data is available. On cache hits, these timestamps differed by at most 1 ms, and were identical 98.8\% of the time. This confirms that cache lookup is instantaneous compared to fetching data from BTrDB.

\subsection{Miss Rate and Miss Penalty}
We identify two metrics that determine the miss rate and miss penalty. First, the latency of requests to BTrDB, including the query processing time in the database, determines the miss penalty. Second, prefetching affects the miss rate and the miss penalty---even if the prefetched results are not available in time, the time spent waiting for them is less, because the request was issued earlier.

In our workload, the user continually moves to new data; therefore, we expect many compulsory misses. However, we still expect the cache hit rate to be nonzero, as the on-screen data may be refreshed from the cache multiple times at the same resolution, in between times where the user crosses the resolution boundary. We quantify the effect of prefetching by measuring the cache miss rate and miss penalty, with and without prefetching enabled.

\subsubsection{Results}\label{sssec:results_reg}
\begin{figure}
    \centering
    \includegraphics[width=0.92\linewidth]{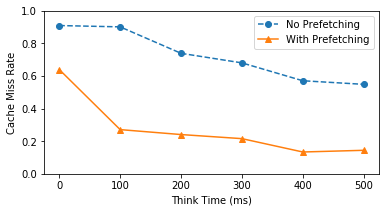}
    \caption{Miss rate, with and without prefetching}
    \label{fig:missrate_reg}
\end{figure}

\begin{figure}
    \centering
    \subfigure[Without prefetching]{
        \includegraphics[width=0.46\linewidth]{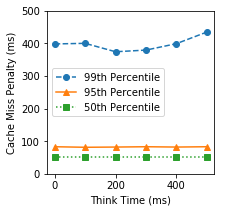}
        \label{subfig:misspenalty_noprefetch_reg}
    }
    \subfigure[With prefetching]{
        \includegraphics[width=0.46\linewidth]{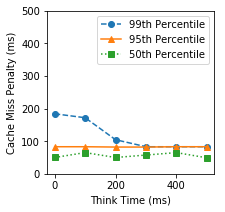}
        \label{subfig:misspenalty_prefetch_reg}
    }
    \caption{Distribution of miss penalty}
    \label{fig:misspenalty_reg}
\end{figure}

We ran the workload described in Section \ref{ssec:workload}, while varying the think time from 0 ms to 500 ms. For each value of the think time, we repeated the experiment 10 times, and measured the latencies of cache lookups in all of the experiments. We also record the fraction of lookups that missed, and required requests to be made to BTrDB, in order to measure the miss rate. We carried out this process twice, once with prefetching enabled, and once with prefetching disabled. The results are shown in Figures \ref{fig:missrate_reg} and \ref{fig:misspenalty_reg}.

Figure \ref{subfig:misspenalty_noprefetch_reg} displays the miss penalty, without any prefetching. \textit{Over 95\% of cache misses are serviced within the interactive threshold of 100 milliseconds.} In other words, even if we use no caching or prefetching at all, the vast majority of requests to BTrDB return at interactive speeds. This is a direct consequence of unifying data reduction techniques in storage and visualization systems. Because \plotter{} draws plots using aggregates that BTrDB has precomputed, it can retrieve data to draw plots at interactive speeds.

That said, the distribution has a long tail. The 99th percentile miss penalty is approximately 400 ms. However, prefetching helps solve this problem. When the think time is at least 300 ms, the 99th percentile miss penalty is less than 100 ms. Furthermore, Figure \ref{fig:missrate_reg} also shows that prefetching is effective at increasing the cache hit rate, especially when the think time is large.

It is expected that prefetching is more effective at reducing the miss penalty when the think time is large, because prefetching only occurs once the data for the current screen has already been fetched. When the think time is very small, the user zooms past the current screen before the data has been fully loaded; therefore prefetching happens less often. Furthermore, even if the prefetching were to happen for small think times, it would be less useful, because the user would move to new data very quickly. It is also expected that the cache miss rate is higher when the think time is small. As the user zooms in, the on-screen data is refreshed every 300 ms during the gesture (see Section \ref{sssec:throttling}), and each refresh counts as a separate miss, if the data is not present. The increased cache miss rate reflects the fact that the user sometimes zooms in before the data is loaded.

\subsubsection{Simulating Non-Ideal Network Conditions}\label{sssec:results_net}
An important challenge addressed by \plotter{} is that, although BTrDB may process queries at interactive speeds, factors such as server load or a flaky network may cause delays in retrieving data. \plotter{}'s client-side data management should ideally provide interactive performance, even under such conditions.

\begin{figure}
    \centering
    \includegraphics[width=0.92\linewidth]{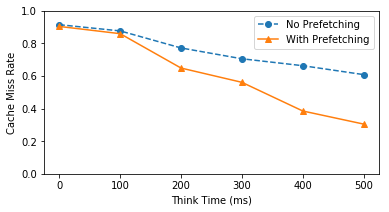}
    \caption{Miss rate on a flaky network}
    \label{fig:missrate_net}
\end{figure}

\begin{figure}
    \centering
    \subfigure[Without prefetching]{
        \includegraphics[width=0.46\linewidth]{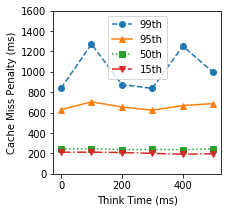}
        \label{subfig:misspenalty_noprefetch_net}
    }
    \subfigure[With prefetching]{
        \includegraphics[width=0.46\linewidth]{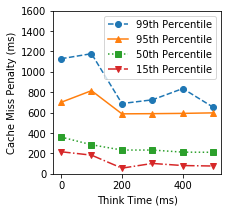}
        \label{subfig:misspenalty_prefetch_net}
    }
    \caption{Distribution of miss penalty on a flaky network}
    \label{fig:misspenalty_net}
\end{figure}

We simulate a flaky network using NetEm. We add a normally distributed delay, with a mean of 200 ms and a jitter of 20 ms, to each outgoing packet, and drop 2\% of outgoing packets. We do not explicitly place a cap on bandwidth, but the variable latency and packet loss interact poorly with TCP, effectively limiting the bandwidth. Using Measurement Lab's speed test hosted by Google, we measured the downlink bandwidth to be approximately 6 Mb/s with the simulated flaky network. Without simulating a flaky network, it is more than 100 Mb/s for the internet connection that we used. We repeated the experiments in Section \ref{sssec:results_reg} while simulating a flaky network. The results are shown in Figures \ref{fig:missrate_net} and \ref{fig:misspenalty_net}.

As expected, the miss penalty is much higher than in Figure \ref{fig:misspenalty_reg}, because the network round trip to fetch data from BTrDB is more expensive. Prefetching reduces the miss penalty when the think time is large. In particular, prefetching allows the miss penalty to drop \textit{below} 200 ms, in cases where a request for data was made earlier due to prefetching. Figure \ref{fig:missrate_net} tells a similar story; when the think time is large, the miss rate is significantly less, because prefetching ensures that necessary data is pre-loaded into the cache.

Prefetching performs slightly worse when the think time is 0 ms.
One explanation is that, when the user moves through data very rapidly, prefetching is ineffective, and only uses bandwidth that could be used to handle normal cache misses.

\subsection{Overview Visualizations}
The latency of querying a full screen of aggregates, where the left and right endpoints of the screen coincide with the earliest and latest data points in the stream, is shown in Table \ref{table:comparison}. We performed this experiment twice, once with the original stream, and again with a derivative stream that repeats the data of the original stream 100 times.
The results confirm that the performance of \plotter{} is essentially independent of the size of the underlying data, allowing it to scale to massive data sets.

To produce this overview visualization, \plotter{} must first make two \textbf{Nearest} queries to find the timestamps for the earliest and latest points in the stream. We do not include the time for these queries in Table \ref{table:comparison}, because overview visualizations do not always require such information (e.g., the overview may show data over the past few hours or days, so the time range does not depend on the stream).
If we had included the time for these queries, the latencies would have been 549 ms / 371 ms and 609 ms / 365 ms.

\section{Related Work}\label{sec:background}
This section compares \plotter{} to related work.

\subsection{Polaris/Tableau (2002, 2008, 2011)}
Tableau is a commercial visualization system for RDBMS systems that grew out of the Polaris~\cite{stolte2002polaris, stolte2008polaris} project. Tableau focuses on multidimensional databases, allowing the user to create a grid of plots, in an organization inspired by Pivot Tables.

Both Tableau and \plotter{} perform data management on the client. \plotter{} uses an in-memory cache; Tableau clients use the Tableau Data Engine~\cite{wesley2011analytic} to store and process data. However, the systems' motivations for client-side data management are different. \plotter{} masks the latency of database queries via its in-memory cache. In contrast, the Tableau Data Engine supports Tableau's \textit{extract} feature, which retrieves a subset of data for offline analysis.

A key difference between Tableau and \plotter{} is that \plotter{} places a greater emphasis on providing a truly interactive interface to explore data, whereas Tableau focuses on providing a flexible, high-quality visualization. The authors of Tableau note that, while the user may change the view and presentation model, the resulting database query may take several tens of seconds~\cite{stolte2002polaris}. A visualization system like Tableau, that makes a query for raw data to produce a visualization, becomes progressively slower as data sets grow, because the time to process a query is linear in the number of records read. In contrast, BTrDB/\plotter{} scales linearly in only the number of aggregates drawn, and can therefore scale to massive data sets.

\subsection{ATLAS (2008)}
ATLAS~\cite{chan2008maintaining} aims to provide an environment for interactive exploration of massive time series. Like \plotter{}, ATLAS identifies the latency of querying large amounts of data as the key obstacle to realizing interactivity, and uses data prefetching as a tool to mask this latency. Furthermore, \plotter{} and ATLAS leverage database technology to improve performance. \plotter{} uses BTrDB, which uses precomputed statistical aggregates to accelerate queries, and ATLAS uses kdb+, which stores data in a column-oriented format.

However, there are important differences between \plotter{} and ATLAS. First, \plotter{} draws the same aggregates that BTrDB can compute efficiently. In contrast, ATLAS' aggregation is computed just in time.
The column-oriented techniques used by kdb+ do not carry through to the plots produced by ATLAS.
Second, as discussed in Section \ref{ssec:automatic_management}, the time to process queries in BTrDB/\plotter{} is essentially independent of the size of raw data bounded by the query, allowing \plotter{} to use a simple prefetching scheme.
Third, \plotter{}, unlike ATLAS, does not place any limits on how fast the user can pan or zoom, and may show perceptually relevant information from the previous view until data for the new view is available. Fourth, \plotter{} is suitable for viewing fast data that is changing, inserted out-of-order, or streaming in.

\subsection{sampleAction (2012)}
The sampleAction~\cite{fisher2012trust} visualization system leverages the idea of incremental databases pioneered by Hellerstein~\cite{hellerstein1997online, hellerstein1999interactive}. This is similar to our approach of unifying data reduction strategies in database and visualization systems.
However, there are important differences. Unlike \plotter{}, which is a visualization tool for timeseries data, sampleAction compares different aggregate queries in a column chart. Furthermore, the focus of \cite{fisher2012trust} is not on the data management problems in building such a system, but rather on the user experience that such a system, if built, would provide.

\subsection{imMens (2013)}
The imMens system~\cite{liu2013immens} is based on the ``principle that scalability should be limited by the chosen resolution of the visualized data, not the number of records,'' which is similar in spirit to \plotter{}. It provides interactive performance for visualizing relational data. The imMens system produces binned plots, and precomputes 3- or 4-dimensional subcubes of the overall data cube on the server. If these subcubes are still large, they are further segmented into multivariate data tiles, which are sent to the client on demand. The client performs query processing on the multivariate data tiles on the GPU, and supports brushing and linking at interactive speeds.

The key difference between imMens and \plotter{} is that imMens focuses on interactive support for brushing and linking\footnote{While \cite{liu2013immens} mentions that the user can pan and zoom, the interface to doing so is not clearly mentioned, and the interactivity of panning and zooming is not evaluated.}, whereas \plotter{} focuses on allowing the user to move interactively through the data. This partially stems from the fact that \plotter{} focuses on visualizing scalar-valued timeseries data, for which brushing and linking is not as relevant; multiple timeseries may be related, but the corresponding points are generally at the same timestamp, making brushing and linking less relevant.

\subsection{ScalaR (2013)}
ScalaR~\cite{battle2013dynamic} is an interactive visualization system for big data. Data are stored in SciDB, and visualized in a web-based front-end written with D3~\cite{bostock2011d3}. The key innovation of ScalaR is an intermediate layer that receives queries from the front-end, requests query plans and metadata from SciDB, and decides whether resolution reduction needs to be performed.
While ScalaR uses data reduction to alleviate bottlenecks in transferring data from the server and managing large query results on the client, it incurs the cost of reading the underlying raw data for each query.

\subsection{M4/VDDA (2014, 2016)}
M4~\cite{jugel2014m4} is a visualization system for timeseries data. Like ScalaR, M4 introduces an intermediate layer between the visualization client and the data store. The intermediate layer transparently rewrites queries to return the first point, last point, lowest point, and highest point in each pixel column, instead of all of the raw data requested by the client. The authors observe that the resulting plot is equivalent, pixel for pixel, to drawing the raw data, but with a smaller data transfer latency and network bandwidth use. The authors also present Visualization-Driven Data Aggregation (VDDA)~\cite{jugel2016vdda}, a generalization of this technique to other types of plots. M4/VDDA has similar pitfalls to ScalaR---while the use of aggregation decreases the data transferred to and rendered by the client, the database query still operates on raw data, and is therefore slow for large data sets. Furthermore, as explained in Section \ref{ssec:plot}, a plot that is pixel-for-pixel equivalent to one containing raw data does nothing to eliminate visual clutter; plotting aggregates such as min, mean, and max, may produce a better visualization.

\subsection{Skydive (2015)}
Perhaps the existing system most similar to \plotter{} is Skydive~\cite{godfrey2015interactive}.
As described in Section \ref{sec:introduction}, \plotter{} improves on Skydive by providing a visualization client that performs automatic data management. In contrast, Skydive requires the user to manually initiate queries by choosing a ``cut'' of data to visualize.

\subsection{Visualization-Aware Sampling (2016)}
Visualization-Aware Sampling (VAS)~\cite{park2016visualization} is a technique to perform data reduction via sampling. Unlike traditional sampling techniques, VAS does not choose a sample randomly; rather, it frames sample selection as an optimization problem, and uses an approximate solution to that problem to select the sample. The key contribution of VAS is that the sample optimizes a visualization-specific metric that ensures that the resulting visualization will be high-quality.

A major difference between VAS and \plotter{} is that VAS uses sampling to perform data reduction. As discussed in Section \ref{sssec:aggregation_discussion}, we prefer aggregation for several reasons.
Furthermore, VAS' optimization function prefers samples that, when plotted, look similar to a plot containing all data points. This has the same pitfall as M4: it does not reduce visual clutter. In contrast, displaying statistical aggregates explicitly may produce a better plot.

\subsection{ASAP (2017)}
Automatic Smoothing for Attention Prioritization (ASAP) \cite{rong2017asap}, is a timeseries visualization operator that computes a moving average over a timeseries, to draw the user's attention to anomalies in fast data~\cite{bailis2017prioritizing, bailis2017macrobase}. The authors present ASAP as an optimization problem, where the window size of the moving average is chosen to minimize the standard deviation of deltas (roughness) while preserving kurtosis (long-term deviations) in the data.

ASAP focuses on producing a visualization that is more useful than simply plotting the raw data. \plotter{}, by displaying min-mean-max aggregates instead of raw data, uses a similar approach. We note that by drawing the min and max aggregates, \plotter{} displays a silhouette of what the plot would look like if every raw point were plotted. The smoothed timeseries that ASAP produces does not convey this information, though the authors note that ASAP could be used in conjunction with a plot of raw data.
We believe that ASAP's goals are complementary to \plotter{}'s; one could use \plotter{}'s interactive exploration in conjunction with ASAP's smoothed timeseries.

\section{Future Work}\label{sec:future}
\plotter{} does not address the problem of moving \textit{across} different time series---for example, starting with many timeseries and finding the one that matches certain criteria.
Algorithms for pattern search in timeseries have been studied.
In TimeSearcher~\cite{hochheiser2004dynamic, buono2005interactive}, the user can draw a timebox on the screen by clicking and dragging the mouse. This filters out any streams that are not contained in the timebox for the timebox's entire horizontal duration. This generalizes to searching for similar streams: many timeboxes can be used, perhaps one per pixel column, where the height of the timeboxes determines the tolerance of the similarity search.
\plotter{} could process timeboxes efficiently: to determine if a stream satisfies a timebox, \plotter{} would check whether the aggregates for that stream, horizontally bounded by the timebox, have ``min'' and ``max'' values within the vertical range of the timebox.
The implementation of timeboxes in \plotter{} is left to future work.

A peculiarity of \plotter{}'s cache organization (see Section \ref{sssec:cache}) is that the aggregates for each resolution are stored in separate tree-based maps, and are treated independently during lookup. Due to the alignment of aggregates, it is possible to compute the aggregates at resolution $x$ (size of aggregates $= 2^x$) from the values of aggregates at resolution $y$ (size of aggregates $= 2^y$) for the same time range, as long as $y < x$. If a query for aggregates at resolution $x$ misses in the cache, but the aggregates at resolution $y$ are present for that time range, it may be more efficient, depending on the value of $x - y$, to compute the aggregates at resolution $x$ on the client's machine, rather than requesting them from BTrDB. We leave the investigation and implementation of this idea to future work.

A related question is whether it is worth spilling the cache to disk, rather than evicting cache entries when the allotted memory is full.
This would be beneficial on a flaky network, where querying BTrDB takes hundreds of milliseconds.
It gives the client an even bigger data management role, similar to the Tableau Data Engine~\cite{wesley2011analytic}. One option is to simply allow the operating system's paging mechanisms to handle the transfer of data from memory to disk; however, existing work suggests that this is unlikely to be successful~\cite{cox1997managing}.
An investigation of this idea is left to future work.

Finally, \plotter{} may benefit from more sophisticated prefetching strategies (e.g., ForeCache~\cite{battle2016dynamic}). We relegate an investigation of this idea to future work.
\section{Conclusion}\label{sec:conclusion}
In order to cope with the growing appetite for data collection and storage, the database community has developed various techniques for working with massive data sets, including databases that quickly return statistical aggregates or compute on samples of data. Meanwhile the visualization community has turned to plotting samples or aggregates to visualize large data sets, in order to solve the problem of ``squeezing a billion points into a million pixels''~\cite{shneiderman2008extreme}.

In this paper, we present \plotter{}, a visualization system for large timeseries data sets. By unifying the data reduction techniques used by databases with those used in visualization, we can process queries with latency proportional to the size of the final visualization, as opposed to the size of the underlying raw data. This makes possible a truly interactive visualization system. The user starts with an overview visualization, and then explores the data by clicking and dragging the plot and zooming in or out with the mousewheel. The user does not explicitly initiate queries. Instead, the visualization client requests data automatically, so that it is available for visualization by the time it is needed.

Existing work has sought to support visualization of big data by moving computation close to the data.
For example, M4 and ScalaR compute the aggregates at the database, to avoid having to transfer all of the raw data to the client. Our work extends this idea by pushing the computation of aggregates even further back, into the data storage itself. That way, only the aggregates, and not the underlying raw data, must be read to process queries.

But our work also represents a fundamental departure from existing designs, which generally decouple the visualization client from the underlying database.
By making the client aware of the storage technology used by the backend database, we create a system that fetches data at interactive time scales, opening the opportunity for automatic data management. This, in turn, introduces new challenges in the client, to manage query results from the server, automatically initiate queries, and cope with ``fast data.''

The core principle of our design---a storage-aware client that supports automatic data management---is applicable beyond timeseries, to other types of data as well. This opens a new direction for research on scalable information visualization systems. We hope that the database and visualization communities will take these ideas even further, to support more effective visual data analytics and bring out the full potential of big data.

%\end{document}  % This is where a 'short' article might terminate

% ensure same length columns on last page (might need two sub-sequent latex runs)
\balance

%ACKNOWLEDGMENTS are optional
\section{Acknowledgments}
This material is based upon work supported by the National Science Foundation Graduate Research Fellowship Program under Grant No.\ DGE-1752814. Any opinions, findings, and conclusions or recommendations expressed in this material are those of the authors and do not necessarily reflect the views of the National Science Foundation. This research is also supported by the U.S. Department of Energy ARPA-E Grant No.\ DE-AR0000340, National Science Foundation Grant No.\ CPS-1239552, Fulbright Scholarship Program, and UC Berkeley Graduate Division (Berkeley Fellowship for Graduate Study).

\bibliographystyle{abbrv}
\bibliography{plotter}  % vldb_sample.bib is the name of the Bibliography in this case

\begin{thebibliography}{10}

\bibitem{agarwal2013blinkdb}
S.~Agarwal, B.~Mozafari, A.~Panda, H.~Milner, S.~Madden, and I.~Stoica.
\newblock Blinkdb: queries with bounded errors and bounded response times on
  very large data.
\newblock In {\em Proceedings of the 8th ACM European Conference on Computer
  Systems}, pages 29--42. ACM, 2013.

\bibitem{ahlberg1994visual}
C.~Ahlberg and B.~Shneiderman.
\newblock Visual information seeking: Tight coupling of dynamic query filters
  with starfield displays.
\newblock In {\em Proceedings of the SIGCHI conference on Human factors in
  computing systems}, pages 313--317. ACM, 1994.

\bibitem{andersen2017package}
M.~P. Andersen.
\newblock Package btrdb.
\newblock \url{https://godoc.org/gopkg.in/btrdb.v4}.
\newblock Online; accessed on 2017-09-02.

\bibitem{andersen2016btrdb}
M.~P. Andersen and D.~E. Culler.
\newblock Btrdb: optimizing storage system design for timeseries processing.
\newblock In {\em Proceedings of the 14th USENIX Conference on File and Storage
  Technologies (FAST 16)}, 2016.

\bibitem{andersen2015distil}
M.~P. Andersen, S.~Kumar, C.~Brooks, A.~von Meier, and D.~E. Culler.
\newblock Distil: Design and implementation of a scalable synchrophasor data
  processing system.
\newblock In {\em Smart Grid Communications (SmartGridComm), 2015 IEEE
  International Conference on}, pages 271--277. IEEE, 2015.

\bibitem{bailis2017macrobase}
P.~Bailis, E.~Gan, S.~Madden, D.~Narayanan, K.~Rong, and S.~Suri.
\newblock Macrobase: Prioritizing attention in fast data.
\newblock In {\em Proceedings of the 2017 ACM International Conference on
  Management of Data}, pages 541--556. ACM, 2017.

\bibitem{bailis2017prioritizing}
P.~Bailis, E.~Gan, K.~Rong, and S.~Suri.
\newblock Prioritizing attention in fast data: Principles and promise.
\newblock CIDR, 2017.

\bibitem{battle2016dynamic}
L.~Battle, R.~Chang, and M.~Stonebraker.
\newblock Dynamic prefetching of data tiles for interactive visualization.
\newblock In {\em Proceedings of the 2016 International Conference on
  Management of Data}, pages 1363--1375. ACM, 2016.

\bibitem{battle2013dynamic}
L.~Battle, M.~Stonebraker, and R.~Chang.
\newblock Dynamic reduction of query result sets for interactive visualizaton.
\newblock In {\em Big Data, 2013 IEEE International Conference on}, pages 1--8.
  IEEE, 2013.

\bibitem{bikakis2017hierarchical}
N.~Bikakis, G.~Papastefanatos, M.~Skourla, and T.~Sellis.
\newblock A hierarchical aggregation framework for efficient multilevel visual
  exploration and analysis.
\newblock {\em Semantic Web}, 8(1):139--179, 2017.

\bibitem{bostock2011d3}
M.~Bostock, V.~Ogievetsky, and J.~Heer.
\newblock D$^3$: data-driven documents.
\newblock {\em IEEE transactions on visualization and computer graphics},
  17(12):2301--2309, 2011.

\bibitem{buevich2013respawn}
M.~Buevich, A.~Wright, R.~Sargent, and A.~Rowe.
\newblock Respawn: A distributed multi-resolution time-series datastore.
\newblock In {\em Real-Time Systems Symposium (RTSS), 2013 IEEE 34th}, pages
  288--297. IEEE, 2013.

\bibitem{buono2005interactive}
P.~Buono, A.~Aris, C.~Plaisant, A.~Khella, B.~Shneiderman, H.~Hochheiser, and
  B.~Schneiderman.
\newblock Interactive pattern search in time series.
\newblock In {\em Proc. SPIE Conference on Visual Data Analytics}, volume 5669,
  pages 175--186, 2005.

\bibitem{chan2008maintaining}
S.-M. Chan, L.~Xiao, J.~Gerth, and P.~Hanrahan.
\newblock Maintaining interactivity while exploring massive time series.
\newblock In {\em Visual Analytics Science and Technology, 2008. VAST'08. IEEE
  Symposium on}, pages 59--66. IEEE, 2008.

\bibitem{cox1997managing}
M.~Cox and D.~Ellsworth.
\newblock Managing big data for scientific visualization.
\newblock In {\em ACM Siggraph}, volume~97, pages 21--38, 1997.

\bibitem{cui2006measuring}
Q.~Cui, M.~Ward, E.~Rundensteiner, and J.~Yang.
\newblock Measuring data abstraction quality in multiresolution visualizations.
\newblock {\em IEEE Transactions on Visualization and Computer Graphics},
  12(5):709--716, 2006.

\bibitem{elmqvist2010hierarchical}
N.~Elmqvist and J.-D. Fekete.
\newblock Hierarchical aggregation for information visualization: Overview,
  techniques, and design guidelines.
\newblock {\em IEEE Transactions on Visualization and Computer Graphics},
  16(3):439--454, 2010.

\bibitem{fisher2012trust}
D.~Fisher, I.~Popov, S.~Drucker, et~al.
\newblock Trust me, i'm partially right: incremental visualization lets
  analysts explore large datasets faster.
\newblock In {\em Proceedings of the SIGCHI Conference on Human Factors in
  Computing Systems}, pages 1673--1682. ACM, 2012.

\bibitem{godfrey2015interactive}
P.~Godfrey, J.~Gryz, P.~Lasek, and N.~Razavi.
\newblock Interactive visualization of big data.
\newblock In {\em Beyond Databases, Architectures and Structures. Advanced
  Technologies for Data Mining and Knowledge Discovery}, pages 3--22. Springer,
  2015.

\bibitem{hellerstein1999interactive}
J.~M. Hellerstein, R.~Avnur, A.~Chou, C.~Hidber, C.~Olston, V.~Raman, T.~Roth,
  and P.~J. Haas.
\newblock Interactive data analysis: The control project.
\newblock {\em Computer}, 32(8):51--59, 1999.

\bibitem{hellerstein1997online}
J.~M. Hellerstein, P.~J. Haas, and H.~J. Wang.
\newblock Online aggregation.
\newblock In {\em ACM SIGMOD Record}, volume~26, pages 171--182. ACM, 1997.

\bibitem{hochheiser2004dynamic}
H.~Hochheiser and B.~Shneiderman.
\newblock Dynamic query tools for time series data sets: timebox widgets for
  interactive exploration.
\newblock {\em Information Visualization}, 3(1):1--18, 2004.

\bibitem{jugel2014m4}
U.~Jugel, Z.~Jerzak, G.~Hackenbroich, and V.~Markl.
\newblock M4: a visualization-oriented time series data aggregation.
\newblock {\em Proceedings of the VLDB Endowment}, 7(10):797--808, 2014.

\bibitem{jugel2016vdda}
U.~Jugel, Z.~Jerzak, G.~Hackenbroich, and V.~Markl.
\newblock Vdda: automatic visualization-driven data aggregation in relational
  databases.
\newblock {\em The VLDB Journal}, 25(1):53--77, 2016.

\bibitem{liu2013immens}
Z.~Liu, B.~Jiang, and J.~Heer.
\newblock immens: Real-time visual querying of big data.
\newblock In {\em Computer Graphics Forum}, volume~32, pages 421--430. Wiley
  Online Library, 2013.

\bibitem{park2016visualization}
Y.~Park, M.~Cafarella, and B.~Mozafari.
\newblock Visualization-aware sampling for very large databases.
\newblock In {\em Data Engineering (ICDE), 2016 IEEE 32nd International
  Conference on}, pages 755--766. IEEE, 2016.

\bibitem{pavlo2009comparison}
A.~Pavlo, E.~Paulson, A.~Rasin, D.~J. Abadi, D.~J. DeWitt, S.~Madden, and
  M.~Stonebraker.
\newblock A comparison of approaches to large-scale data analysis.
\newblock In {\em Proceedings of the 2009 ACM SIGMOD International Conference
  on Management of data}, pages 165--178. ACM, 2009.

\bibitem{rafiei2005effectively}
D.~Rafiei.
\newblock Effectively visualizing large networks through sampling.
\newblock In {\em Visualization, 2005. VIS 05. IEEE}, pages 375--382. IEEE,
  2005.

\bibitem{rong2017asap}
K.~Rong and P.~Bailis.
\newblock Asap: Prioritizing attention via time series smoothing.
\newblock {\em Proceedings of the VLDB Endowment}, 10(11), 2017.

\bibitem{shneiderman1994dynamic}
B.~Shneiderman.
\newblock Dynamic queries for visual information seeking.
\newblock {\em IEEE software}, 11(6):70--77, 1994.

\bibitem{shneiderman1996eyes}
B.~Shneiderman.
\newblock The eyes have it: A task by data type taxonomy for information
  visualizations.
\newblock In {\em Visual Languages, 1996. Proceedings., IEEE Symposium on},
  pages 336--343. IEEE, 1996.

\bibitem{shneiderman2008extreme}
B.~Shneiderman.
\newblock Extreme visualization: squeezing a billion records into a million
  pixels.
\newblock In {\em Proceedings of the 2008 ACM SIGMOD international conference
  on Management of data}, pages 3--12. ACM, 2008.

\bibitem{stoica2013big}
I.~Stoica.
\newblock For big data, moore's law means better decisions.
\newblock
  \url{https://amplab.cs.berkeley.edu/for-big-data-moores-law-means-better-decisions/}.
\newblock Online; accessed on 2017-09-07.

\bibitem{stolte2002polaris}
C.~Stolte, D.~Tang, and P.~Hanrahan.
\newblock Polaris: A system for query, analysis, and visualization of
  multidimensional relational databases.
\newblock {\em IEEE Transactions on Visualization and Computer Graphics},
  8(1):52--65, 2002.

\bibitem{stolte2008polaris}
C.~Stolte, D.~Tang, and P.~Hanrahan.
\newblock Polaris: a system for query, analysis, and visualization of
  multidimensional databases.
\newblock {\em Communications of the ACM}, 51(11):75--84, 2008.

\bibitem{stonebraker2005c}
M.~Stonebraker, D.~J. Abadi, A.~Batkin, X.~Chen, M.~Cherniack, M.~Ferreira,
  E.~Lau, A.~Lin, S.~Madden, E.~O'Neil, et~al.
\newblock C-store: a column-oriented dbms.
\newblock In {\em Proceedings of the 31st international conference on Very
  large data bases}, pages 553--564. VLDB Endowment, 2005.

\bibitem{wesley2011analytic}
R.~Wesley, M.~Eldridge, and P.~T. Terlecki.
\newblock An analytic data engine for visualization in tableau.
\newblock In {\em Proceedings of the 2011 ACM SIGMOD International Conference
  on Management of data}, pages 1185--1194. ACM, 2011.

\bibitem{zaharia2010spark}
M.~Zaharia, M.~Chowdhury, M.~J. Franklin, S.~Shenker, and I.~Stoica.
\newblock Spark: Cluster computing with working sets.
\newblock {\em HotCloud}, 10(10-10):95, 2010.

\end{thebibliography}

%\subsection{References}
%Generated by bibtex from your ~.bib file.  Run latex,
%then bibtex, then latex twice (to resolve references).

%APPENDIX is optional.
% ****************** APPENDIX **************************************
% Example of an appendix; typically would start on a new page
%pagebreak

%\begin{appendix}
%You can use an appendix for optional proofs or details of your evaluation which are not absolutely necessary to the core understanding of your paper.

%\section{Final Thoughts on Good Layout}
%Please use readable font sizes in the figures and graphs. Avoid tempering with the correct border values, and the spacing (and format) of both text and captions of the PVLDB format (e.g. captions are bold).

%At the end, please check for an overall pleasant layout, e.g. by ensuring a readable and logical positioning of any floating figures and tables. Please also check for any line overflows, which are only allowed in extraordinary circumstances (such as wide formulas or URLs where a line wrap would be counterintuitive).

%Use the \texttt{balance} package together with a \texttt{\char'134 balance} command at the end of your document to ensure that the last page has balanced (i.e. same length) columns.

%\end{appendix}

\end{document}